\documentclass[12pt,preprint]{aastex}

\def\lea{\mathrel{<\kern-1.0em\lower0.9ex\hbox{$\sim$}}}
\def\gea{\mathrel{>\kern-1.0em\lower0.9ex\hbox{$\sim$}}}
\def\Msun{\hbox{M$_{\odot}$}}

\slugcomment{Accepted by The Astrophysical Journal}

\shorttitle{New Clusters in M33}
\shortauthors{San Roman et al.}

\begin{document}


\title{Newly Identified Star Clusters in M33. II. \\ Radial HST/ACS Fields
\footnote{Based on observations made with the NASA/ESA Hubble Space 
Telescope, obtained from the data archive at the Space Telescope Science 
Institute. STScI is operated by the Association of Universities for Research in 
Astronomy, Inc. under NASA contract NAS 5-26555.}}

\author{Izaskun San Roman and Ata Sarajedini}
\affil{Department of Astronomy, University of Florida, 211 Bryant Space
Science Center, Gainesville, FL 32611-2055}
\email{izaskun@astro.ufl.edu, ata@astro.ufl.edu}

\author{Donald R. Garnett}
\affil{801 W. Wheatridge Drive, Tucson AZ 85704}
\email{drgarnett@msn.com}

\author{Jon A. Holtzman}
\affil{New Mexico State University, Las Cruces, NM}
\email{holtz@nmsu.edu}



\begin{abstract}
We present integrated photometry and color-magnitude diagrams 
for 161 star clusters in M33, of which 115 were previously uncataloged, 
using the Advanced Camera For Surveys Wide Field Channel 
onboard the Hubble Space Telescope. The integrated V-band
magnitudes of these clusters range from $M_V$$\sim$--9 to as
faint as $M_V$$\sim$--4, extending the depth of
the existing M33 cluster catalogs by $\sim$1 mag.  
Comparisons of theoretical isochrones to the color-magnitude
diagrams using the Padova models yield ages for 148 of these star clusters. 
The ages range from Log (t)$\sim$7.0 to Log (t)$\sim$9.0. Our color-magnitude
diagrams are not sensitive to clusters older than $\sim$1 Gyr.
We find that the variation of the clusters' integrated colors and absolute 
magnitudes with age is consistent with the predictions of simple stellar 
population models. These same models suggest that the masses of
the clusters in our sample range from $5\times10^3$ to $5\times10^4$ \Msun.
\end{abstract}



\section{Introduction}

Star clusters provide a unique and powerful probe useful for studying 
the star formation histories of galaxies. In particular, 
the ages, metallicities and kinematics of 
star clusters bear the imprint of the galaxy formation process. These correlations 
underscore the importance of compiling complete catalogs of clusters in 
nearby galaxies. At a distance of ~870kpc (distance modulus=24.69; \citet{Galleti2004}), 
M33 is the only nearby late-type spiral galaxy and it 
provides a notable connection between the cluster  populations of 
earlier-type spirals and the numerous, nearby later-type dwarf galaxies.

There have been a number of M33 cluster catalogs published since
the pioneering work of Hiltner (1960). The reader is referred to the work of
\citet{Sarajedini2007} (hereafter referred to as SM), which merged
all of the modern catalogs compiled before 2007, for a summary of the
properties of all of these catalogs. 
Even the results of Park \& Lee (2007), which appeared after the 
publication of SM have been incorporated into the web-based
version of the SM catalog 
\footnote{http://www.astro.ufl.edu/$\sim$ata/cgi-bin/m33\_cluster\_catalog/index.cgi}.
This catalog contains 479 candidates of which 264 are confirmed clusters 
based on $\textit{HST}$ and high resolution ground-based imaging. 
The most recent work in this field is that of 
\cite{Zlocz2008}(hereafter ZKH), which presents a catalog of 4780 extended 
sources in a 1 square degree region around M33 including 3554 new 
candidate stellar clusters. For the purpose of the present paper,
it is important to take special note of the two cluster catalogs by 
Chandar et al. (1999, hereafter CBF99; 2001, hereafter CBF01,
or collectively as CBF) and the one by Sarajedini et al (2006) 
because we will adopt similar reduction and analysis 
techniques as these previous studies.

In general, ground-based imaging cannot clearly distinguish a star
cluster from another type of extended source (e.g. nebula, galaxy),
but Hubble Space Telescope (HST) imaging with the Wide-Field Planetary Camera 2
(WFPC2) or the Advanced Camera for Surveys (ACS) provides the spatial resolution
necessary to provide an unequivocal determination. It is for this 
reason that we have undertaken the present study. In addition,
and just as importantly, the HST observations allow us to construct
color-magnitude diagrams (CMDs) of the star clusters, which 
can be used to estimate ages and investigate any 
correlations that might exist between cluster ages and integrated 
photometric properties as well as allowing us to determine star cluster masses. 
Section 2 describes the observations and data reduction, and \S 3 discusses 
the identification and integrated magnitudes of the clusters. 
\S 4 includes a statistical analysis of the most recent previous
survey of M33 clusters by \cite{Zlocz2008}. The analysis of 
the CMDs is presented in \S 5 while an analysis of the star cluster properties 
is in \S 6. Lastly, \S 7 presents our conclusions.



\section{Observations and Data Reduction}

\begin{figure}
\epsscale{0.9}
\plotone{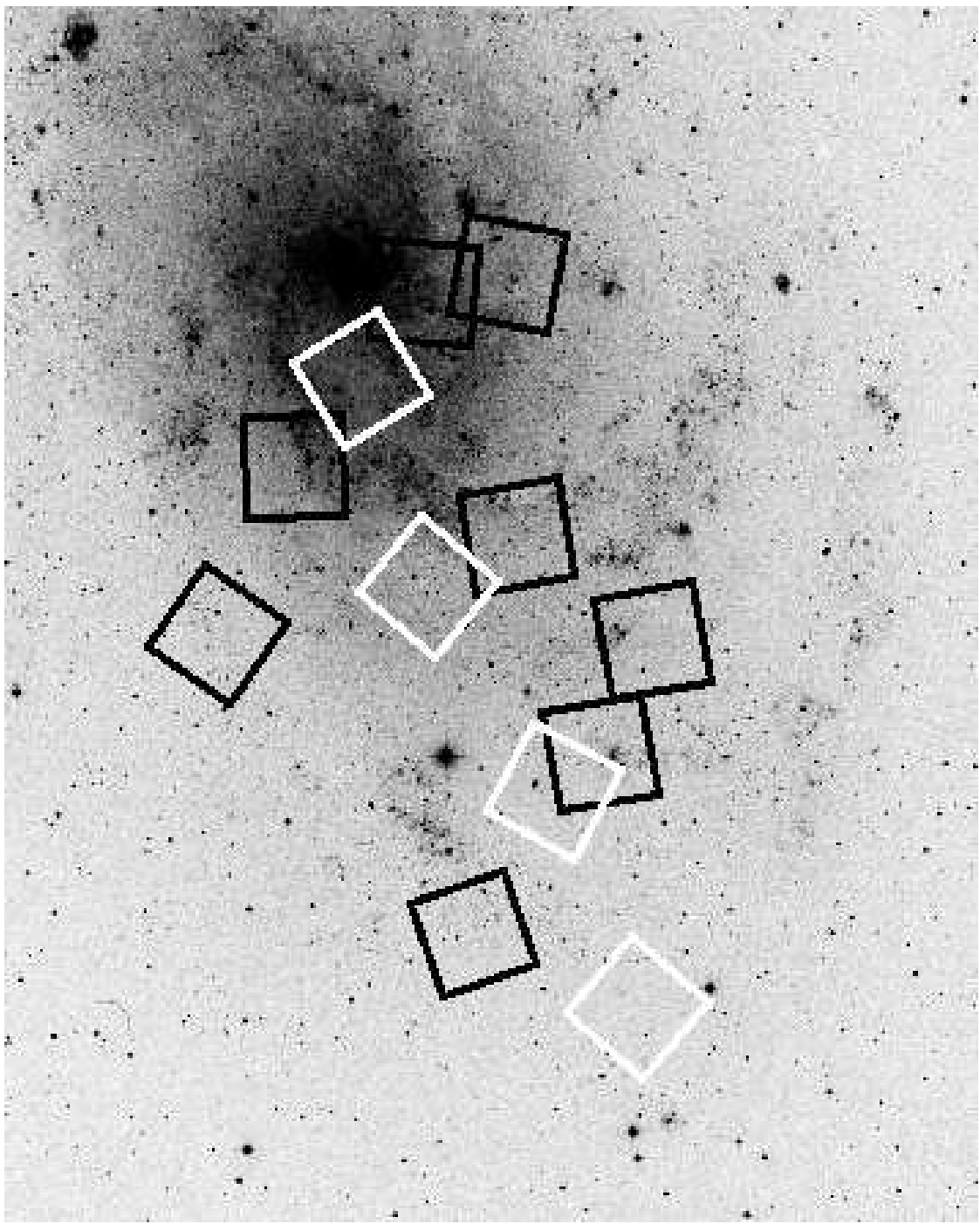}
\caption{ Location of our observed ACS/WFC fields overplotted on an image of M33. North is up and east is to the left. White squares are primary fields and black squares are parallel fields.}
\label{fields}
\end{figure}

The observations for the present study were obtained with the Advanced Camera
for Surveys Wide Field Channel ($\textit{ACS/WFC}$) onboard the Hubble Space
Telescope ($\textit{HST}$). With a pixel scale of $0.049 ''$ $pixel^{-1}$ and 
a field of view of 3.3' x 3.3', $\textit{ACS/WFC}$ is able to resolve
individual stars at the distance of M33.
Twelve $\textit{HST/ACS}$ fields from the GO-10190 
program (PI: D. Garnett) have been analyzed. Four primary fields were obtained
along the major axis of M33. Eight coordinated parallel fields were also secured
along both sides of the primary-field axis. 
Figure \ref{fields} shows the locations of these fields. We have three filters 
for the primary observations (F475W, F606W, F814W) and two filters for the 
parallel  images (F606W, F814W). Table \ref{observations} presents a 
summary of the observations.

All of the images were calibrated through the standard pipeline-process and
downloaded from the $\textit{HST}$ archive. The `FLT' 
images (Table \ref{observations}), which were used for the
point source photometry, were first multiplied by the geometric correction
image to correct for the fact that each ACS pixel subtends a different angle on
the sky. Then, the data quality files were applied by setting the values of bad
pixels to a large negative number so that the photometry software will ignore them.
The photometry was performed using the DAOPHOT/ALLSTAR and ALLFRAME routines 
(Stetson 1994) following the same procedure as that used by 
\cite{Sarajedini2000}. A detailed description of how the point spread functions 
(PSFs) were constructed has been presented in \cite{Sarajedini2006}. 
The corresponding frames in the F606W and F814W filters were matched to obtain mean 
instrumental magnitudes of common stars which were then matched to form colors. 
The photometry has been corrected for the charge transfer efficiency (CTE) 
problem that ACS suffers using the prescription of \cite{Riess2004}. 
In addition, the  
theoretical transformation of \citet{Sirianni2005} was used in order to convert 
the magnitudes to the ground-based Johnson-Cousins system.

The pipeline-processed drizzled (`DRZ') images were used for the integrated cluster 
photometry and were similarly obtained from the \textit{HST} archive. In fields where
multiple drizzled images were produced by the pipeline, we derived positional offsets 
between these frames using the \textit{imshift} and \textit{imcombine} 
tasks in IRAF to allow us to produce one master image per filter per field. 
The drizzling process removes the sky background and corrects the counts
to an exposure time of one second. As a result,  to make the
calculation of photometric errors more straightforward, each   
DRZ image was multiplied by the exposure time and the background sky value
was added back before performing photometry on these frames. 
The resultant images were then used 
in the derivation of the integrated cluster photometry.

\section{Cluster Identification and Photometry}

\begin{figure}
\epsscale{0.9}
\centering
\includegraphics[width=0.9\textwidth]{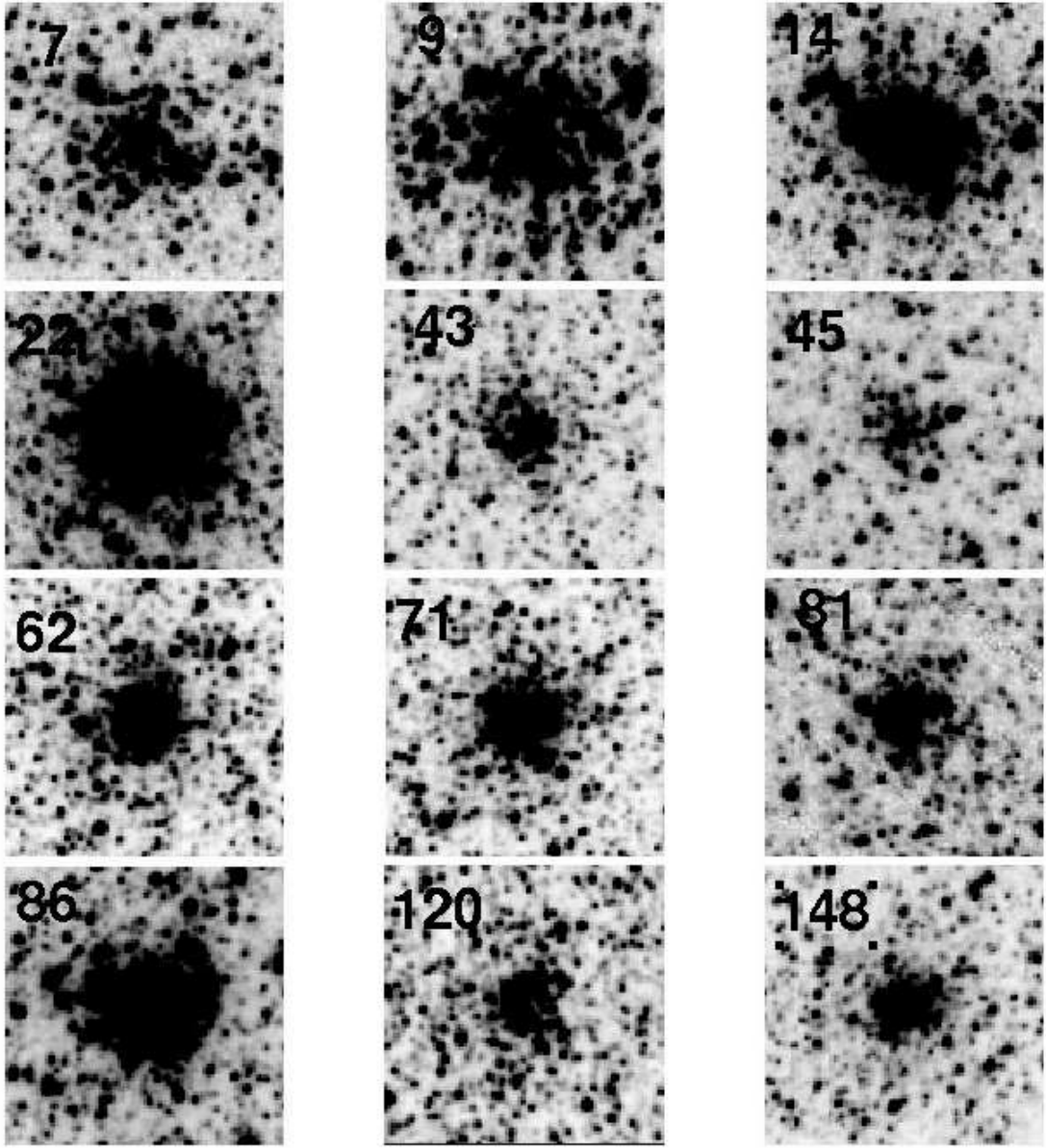}
\caption{Representative sample of star clusters present in our fields in the F606W filter. Each image is shown with the same gray-scale intensity and 5'' arsec on a side, with north up and east to the left.}
\label{mosaic}
\end{figure}

Star clusters are easily resolved on the \textit{HST/ACS} images so the 
selection of objects has been done by visual inspection of each image. 
This is the technique employed in several previous papers where HST imaging
was used (CBF, Sarajedini et al. 2006).
Figure \ref{mosaic} shows sample images of some of the 161 clusters identified in
this study. The cluster positions have been determined by convolving
each image with an elliptical gaussian of $\sigma$=10 pixels. This kernel size
was chosen so that the convolution process would yield a smooth cluster
profile that is conducive to the next step, which is the application of
the IRAF \textit{imexamine} task. This routine was applied to the smoothed
cluster profiles to determine the cluster centers. The optimum 
pixel coordinate positions were transformed to right ascension and 
declination using the World Coordinate System in the image headers.
Based on the work of SM, which compared the positional accuracy
of several catalogs with those determined using the Local Group Survey
images of Massey et al. (2006), we estimate an internal precision of
approximately $\pm$0.1 arcsec and an absolute accuracy of about
$\pm$0.4 arcsec for these quoted cluster positions. 

The integrated magnitudes and colors of each cluster have been calculated 
using the aperture photometry routines in DAOPHOT (Stetson 1987). To be 
consistent with the previous work of Sarajedini et al. (2006),
we have adopted an aperture radius of 2.2'' for the magnitude 
measurements and 1.5" for the colors. 
Note that CBF used the same aperture
size for their magnitudes but a variable aperture ranging from 1.0" to 2.2"
within which to measure cluster colors.
Like in the CBF study, the background sky is always determined in an annulus with 
an inner radius of 3.5'' and an outer radius of 5.0''. Once again, these magnitudes 
have been corrected for CTE based on \cite{Riess2004} and calibrated to
the ground-based system using the synthetic transformations of \cite{Sirianni2005}.

Table \ref{table} details the position of each cluster as well as its V 
magnitude, B--V color, V--I color, reddening, age, mass and also, if applicable, 
the alternative identification in \cite{Sarajedini2007}. We could not determine the 
integrated magnitudes for two of the clusters
because of their location near the edge of the field. 
The formal random errors on 
the magnitudes and colors are all less than 0.01 mag because of the high 
signal-to-noise 
ratio of these clusters. Cluster candidates number 139 and 59 in the catalog of 
\cite{Sarajedini2007} 
have been rejected as clusters in this study based on visual inspection. 

\subsection{Comparison with Previous Photometry}

\begin{figure}
\begin{center}
\epsscale{0.9}
\includegraphics[]{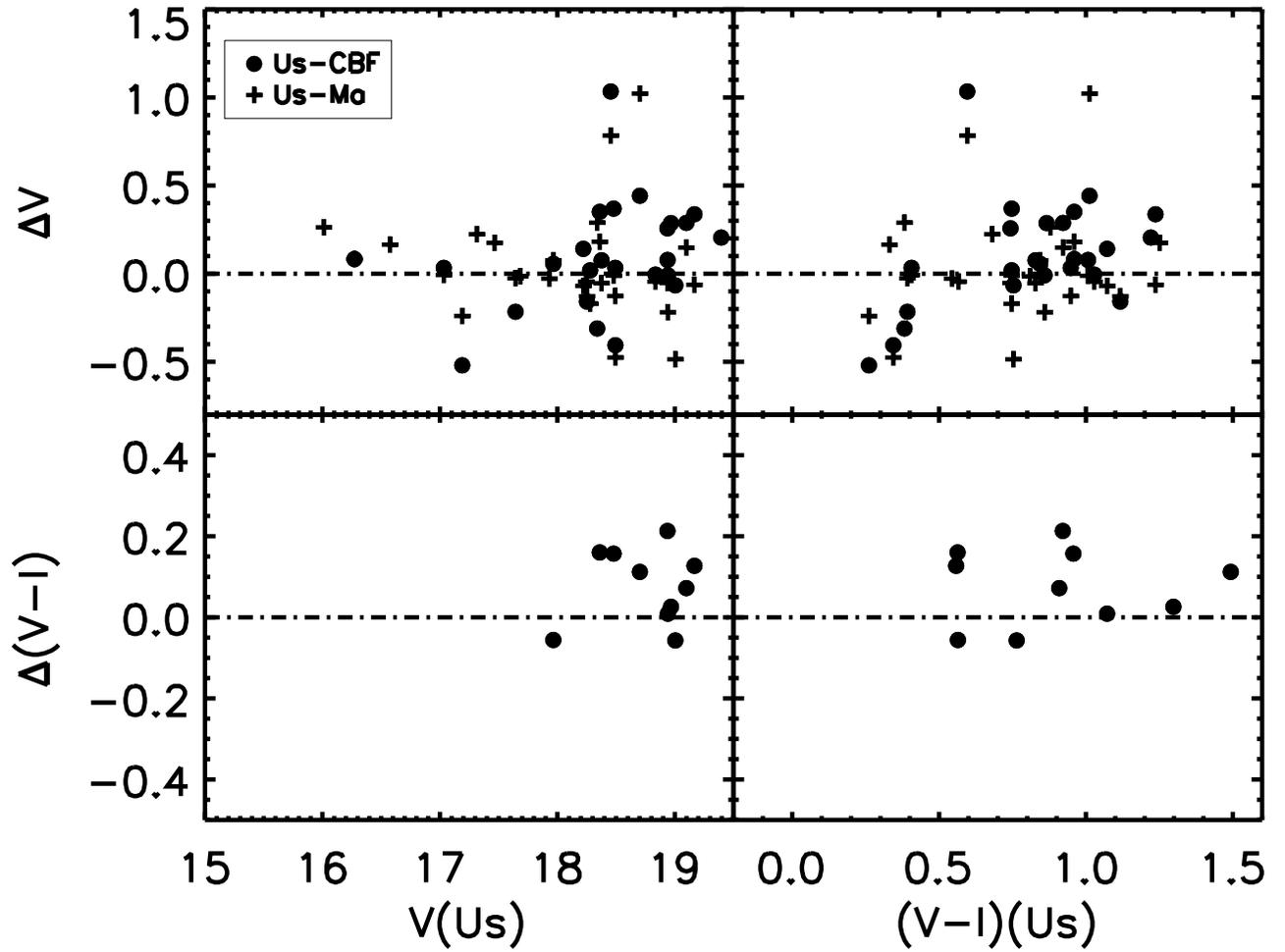}
\caption{Comparison of the integrated cluster photometry from the present
study with that of CBF and Ma et al. (2001; 2002a; 2002b).}
\label{comparison}
\end{center}
\end{figure}

Comparing our results with the CBF study, we find a mean difference of 
$\langle$$\Delta$V(Us-CBF)$\rangle$= 0.09 $\pm$ 0.06 and 
$\langle$$\Delta$(V-I)$\rangle$= 0.07 $\pm$ 0.03 while comparing with 
Ma et al. (2001; 2002a; 2002b), we arrive at a difference of 
$\langle$$\Delta$V(Us-Ma)$\rangle$= 0.04 $\pm$ 0.05, 
where the uncertainties are standard errors of the mean. As we shall see
below, these photometric differences are not unexpected for integrated
photometry of extended objects such as star clusters (see also Table 2 of
Sarajedini \& Mancone 2007).
Fig. \ref{comparison} shows the offset in magnitude and color in both cases. The 
integrated B magnitudes are not plotted due to a lack of significant numbers of 
clusters with which to compare. 

In order to examine these results in more detail, we analyzed the 
magnitude and color offsets as a function of field and position. Not 
surprisingly, the mean difference is larger in fields and positions near the 
center of the galaxy as a result of the higher degree of crowding. 
Analysis of the distance distribution of our sample reveals that 50\% of the
CBF clusters with measured V magnitudes are inside a distance of 1.7 kpc
from the center of M33. The mean magnitude difference for the clusters inside 
this region is $\langle$$\Delta$V(Us-CBF)$\rangle$= 0.17 $\pm$ 0.07 while for 
those outside this region, the difference is 
$\langle$$\Delta$V(Us-CBF)$\rangle$= 0.02 $\pm$ 0.09. Examining different distance
ranges, moving progressively outward to include more CBF clusters, we find 
the same tendency where clusters in the inner (more crowded) regions
display a larger mean magnitude difference as compared with the outer (less
crowded) regions.

It is important to note that we also performed additional tests of the photometry
to investigate the effects of spatial resolution and errors in cluster centering.
The CBF photometry comes mostly from the 
Wide Field (WF) CCDs that are part of
the WFPC2 onboard HST. The spatial resolution
of the WF CCDs is roughly 4 times coarser as compared with ACS/WFC. We performed
photometry of a subset of our clusters using ACS/WFC images that were resampled
to replicate the resolution of the WF CCDs. We found no significant difference
between these results and the magnitudes as measured on the original ACS/WFC
frames.
Additionally, we analyzed the sensitivity of our photometry to 
the adopted cluster center by using the values from the \cite{Sarajedini2007} 
catalog, which are measured from ground-based images taken from the 
Local Group Survey using the MOSAIC instrument as well as from the work of
CBF. Again, we find no significant difference in the magnitudes and colors
of the clusters we have measured. 

A further check of our photometry is provided by the realization that,
as shown in Fig. \ref{fields}, several of our fields exhibit
significant overlap. There are 10 clusters that have multiple measurements of
their magnitudes and colors: one common cluster between d2 and f4, five
between f1 and f2 
and four between d3 and f7. Comparing the photometry of  these common 
clusters, we 
obtain an offset in V magnitude of $\Delta$V=0.042 $\pm$ 0.026 and an offset in 
color of $\Delta$(V--I)=0.060 $\pm$ 0.012. 

At first glance, this level of disagreement between measurements of the 
same clusters in different 
fields seems to be a potential cause for concern. However, 
an examination of a similar situation encountered by CBF99 and CBF01
reveals that their studies show offsets of the same order. In particular, as pointed
out by SM, there are 3 clusters in common between CBF99 and CBF01,
and these objects appear on two different WFPC2 fields. The mean
difference in the V mags of these clusters is 
$\Delta$V(CBF99-CBF01) = 0.106 $\pm$ 0.048. 

\subsection{Comparison with Previous Catalogs}

\begin{figure}
\begin{center}
\epsscale{0.9}
\includegraphics[]{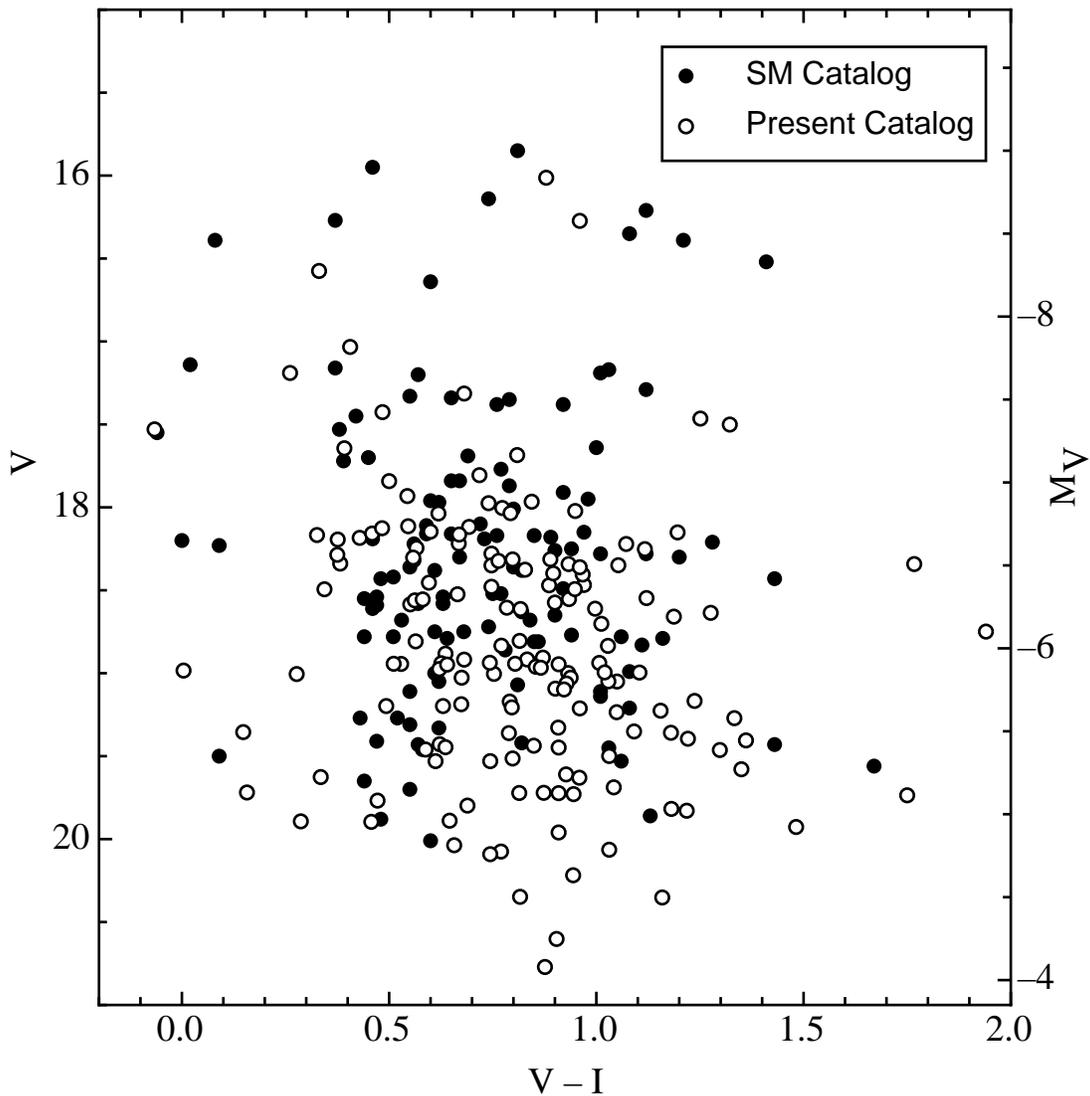}
\caption{Color-magnitude diagram of the genuine M33 clusters from the
SM catalog (open circles) as compared with those from the present study
(filled circles). The distribution of the two types of points is largely similar
except that the present catalog contains more faint star clusters than that
of SM. The faintest clusters here have $M_V$$\sim$--4.0 rivaling 
the faintest globular clusters in the Milky Way and populous clusters
in the LMC.}
\label{CMD}
\end{center}
\end{figure}

As noted in \S 1, the SM catalog is the pre-eminent compilation containing all
of the existing information for M33 star clusters up to 2007. Of the 161 clusters 
identified in the present work, 46 have been previously cataloged in 
\cite{Sarajedini2007}. Fig. \ref{CMD} shows the
integrated magnitude CMD for all of the genuine clusters in the SM catalog.
We see that the clusters span a magnitude range of 16 $\lea$V$\lea$20,
which corresponds to --9$\lea$$M_V$$\lea$--5, using our adopted distance
modulus of $(m-M)_{0}$=24.69 \citep{Galleti2004}. Fig. \ref{CMD} also
illustrates the locations of the clusters identified in the present study. 
This new sample extends to $M_V$$\sim$--4, which is $\sim$1 mag
fainter than the least luminous clusters in the SM catalog, closer
to the faintest clusters in the Large Magellanic Cloud 
\citep{Sarajedini2007}. 

We can also compare our cluster catalog with the most recent catalog produced
by \cite{Zlocz2008}(hereafter ZKH); they present a photometric survey for stellar 
clusters in M33 based on deep ground-based images obtained with the MegaCam instrument 
on the Canada-France-Hawaii telescope, and their classifications are based on visual inspection 
of the images. Their catalog contains 4780 extended sources in a region approximately
1 x 1 degree. Among these, there are 3554 new candidate stellar clusters of 
which 122 are relatively bright, likely globular clusters. 

Sixty of the 3554 objects are present in our HST/ACS fields.  The ZKH catalog 
suggests that 51 of these are new candidate stellar
clusters; however our study shows that only 21 objects resolve into stars and
therefore appear to be clusters. Some of the other objects that ZKH consider to
be extended turn out to be close groupings of two or three stars. Others appear
to be background galaxies or diffuse nebulae. Table 3 provides a 
cross-identification of the 21 common objects that are genuine clusters.
If we extrapolate the ratio
of true clusters to total cluster candidates in the ZKH catalog, this suggests
 that only around
40 $\%$  of the 3554 proposed candidates from ZKH will be actual stellar
clusters.  Nonetheless, even if there are $\sim$1400 stellar clusters in M33,
it would have the highest number of known star clusters per unit luminosity of 
any spiral galaxy in the Local Group.  

\section{Color-Magnitude Diagrams}

\begin{figure}
\begin{center}
\epsscale{0.9}
\includegraphics[bb= 59 409 622 1038,width=0.41\textwidth,clip=true]{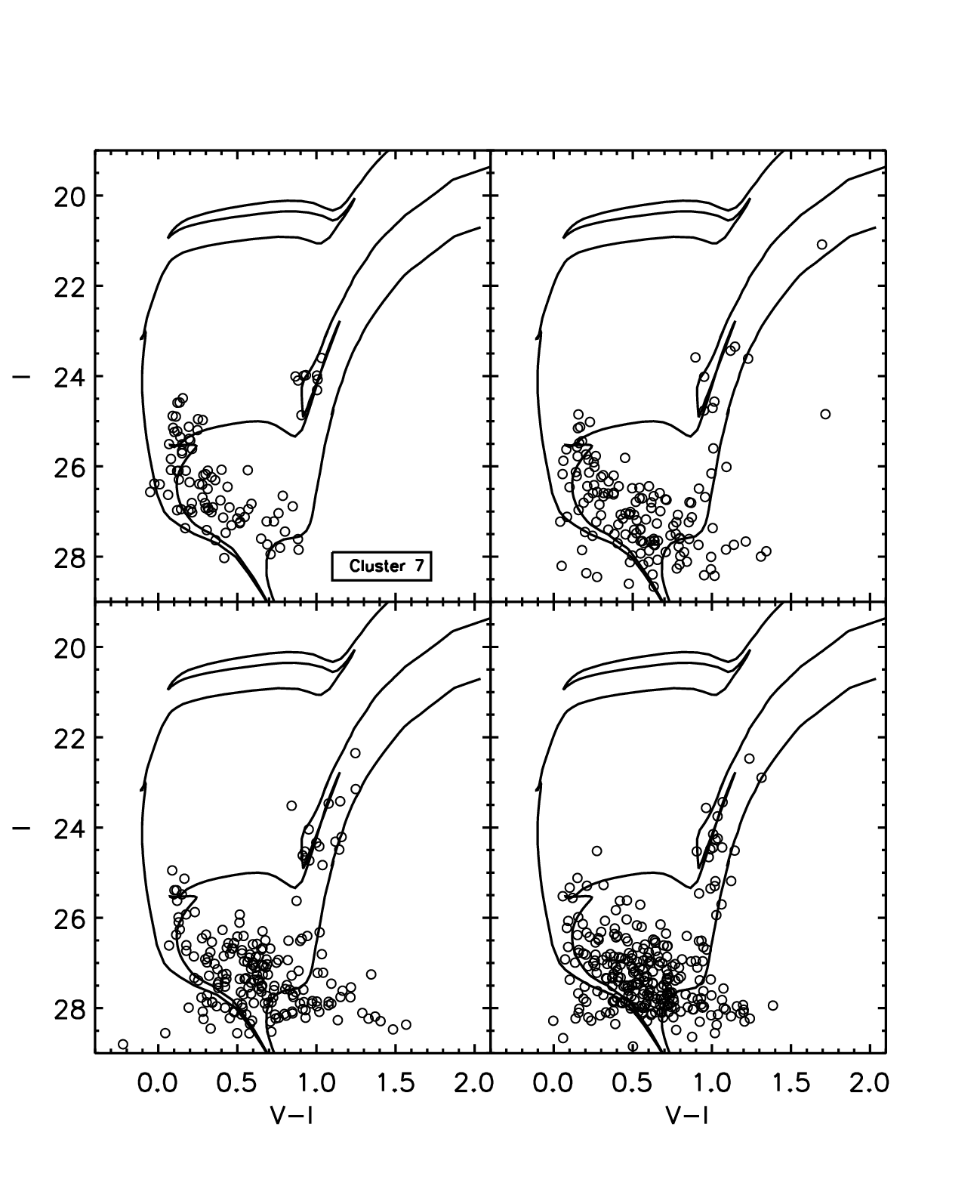}
\hfil
\includegraphics[bb= 87 377 660 1110,width=0.41\textwidth,clip=true]{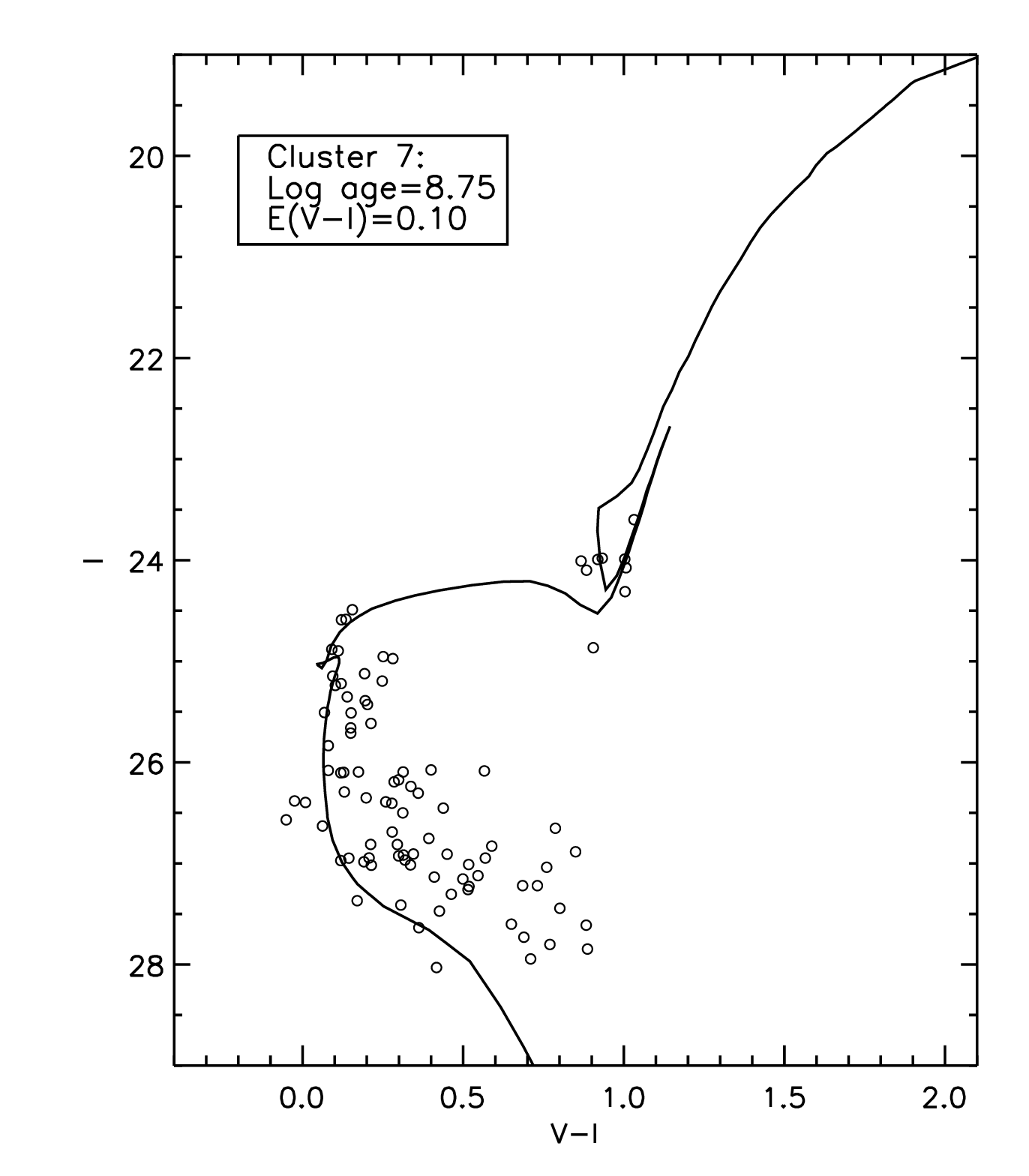}
\caption{$\textit{Left}$: CMD for the region around cluster 7. The top left panel shows the stars within 1'' of the cluster center, the top right panel includes the stars between 1'' and 2'', while the stars between 2'' and 3'' and 3'' and 4'' are displayed in the bottom right panel and the bottom left, respectively. The solid lines correspond to isochrones from \cite{Girardi2000} for ages of $10^8$, $10^9$ and $10^{10}$ yr and a metallicity of Z=0.004. $\textit{Right}$: The best fit isochrone for stars within 1" of the cluster center.}
\label{cmd1}
\end{center}
\end{figure}

Intrinsic properties such as age, metallicity and reddening govern the 
integrated magnitudes and colors of clusters. Using the PSF-photometry of 
each star in each field we are able to examine these parameters by 
constructing CMDs of each cluster. Figs. \ref{cmd1} - \ref{cmd5} show 
radial CMDs and the best fit isochrone for the region around each cluster for seven 
clusters in our list as an example of the procedure we followed and the range of ages found. 
The top-left to the bottom-right in the left panel correspond to stars within 1'' of 
the cluster center, between 1'' and 2'', 2'' and 3'' and finally 
between 3'' and 4''. The solid lines represent theoretical isochrones 
from \cite{Girardi2000} for ages of $10^8$, $10^9$ and $10^{10}$ yrs 
with a metallicity of Z=0.004. This metal abundance has been chosen as a 
representative mean of the disk abundance gradient based on the work
of \cite{Kim2002} (see also Sarajedini et al. 2006). However, we point out that for ages younger than
$\sim$1 Gyr, there is very little sensitivity of the isochrone-derived
age on the assumed metal abundance.
The isochrones have been shifted by a distance modulus of 
$(m-M)_{0}$=24.69 \citep{Galleti2004} and a line-of-sight
reddening value of E(V-I)=0.06 \citep{Sarajedini2000}. Then, 
we overplot the isochrones in the observed CMDs looking for 
the best fit to the main sequence turnoff (MSTO) region. 
Based on \citet{Cardelli1989}, we adopt the following relation
between the extinction and the reddening: 
$A_{V}=1.3$$\cdot$$E(V-I)$.

The left panel illustrates the gradual decrease in cluster stars at 
increasing distance from the cluster center. Consequently, CMDs within 1'' 
of the cluster center display a significant fraction of stars belonging to the cluster 
revealing features to estimate ages, while the lower panels have been used 
to monitor non-cluster stars. Right panel in each figure shows the best fit isochrone for stars within 1" of the cluster center.
In this way, the ages of 148 star clusters 
have been estimated from a comparison of the main-sequence turnoff photometry 
with theoretical isochrones. In some cases, an additional reddening adjustment
was needed in order to align the main sequence of the isochrones with the data. 
In several instances, the main sequence and the turnoff are not satisfactorily 
defined to permit a comparison with isochrones. We estimate a precision in the 
isochrone-fitting ages of $\pm$0.05 dex based on neighboring 
isochrones (in age) that could also potentially fit the data.
In the case of clusters 12, 15, 18 and 111, the isochrone 
fit is not as well established and the precision of the age reaches 
0.1 dex. The precision of the 
reddenings is approximately $\pm$ 0.05mag. The ages and reddenings obtained 
during this process are listed in Table \ref{table}. 


\section{Analysis}

\begin{figure}
\begin{center}
\epsscale{0.9}
\includegraphics[]{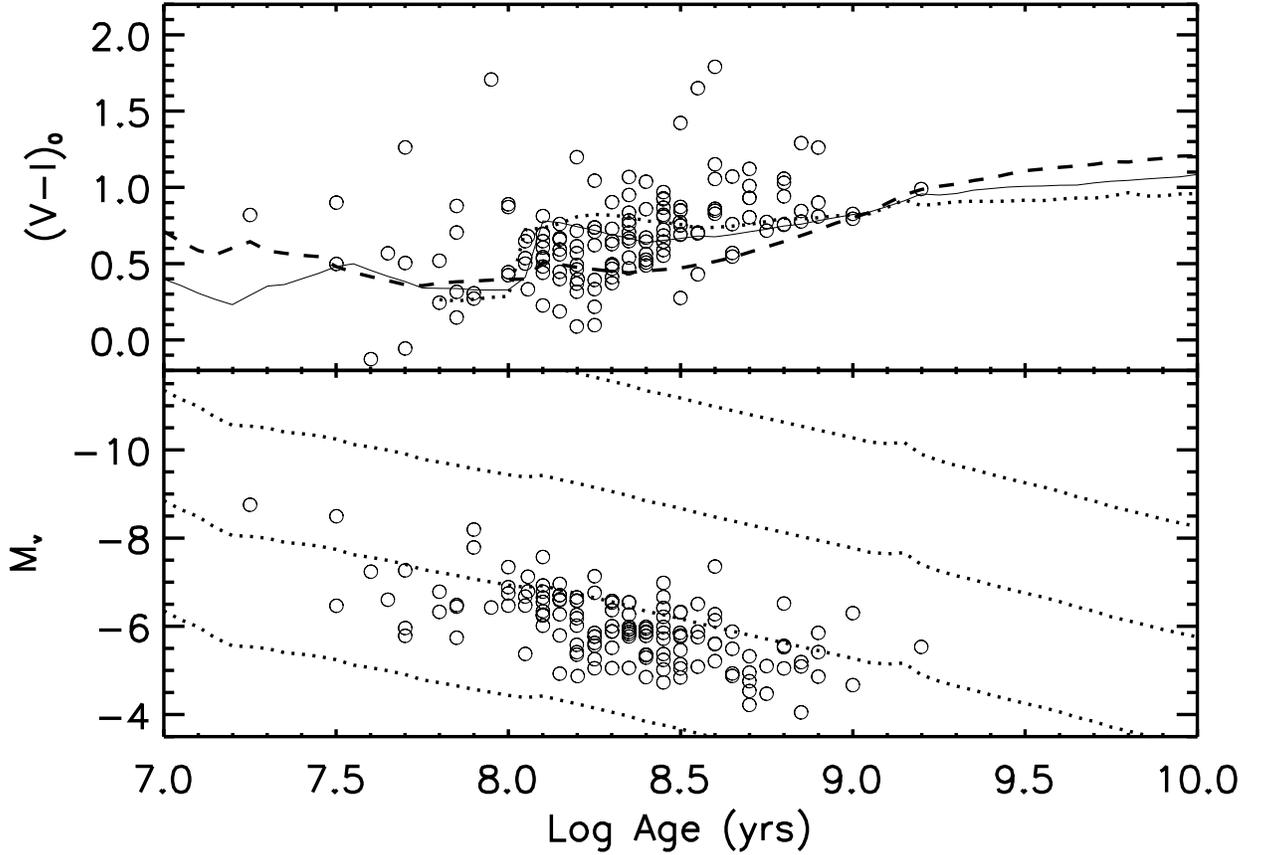}
\caption{The top panel shows the variation of the clusters' integrated 
(V--I)$_0$ colors with age. The correction for reddening has been effected
using the reddening values obtained in the isochrones fitting process. The lines represent the expected relations for a simple stellar population from \citet{Girardi2002} with Z=0.004 (solid line), Z=0.001(dotted line) and Z=0.019 (dashed line). The 
lower panel shows the extinction corrected absolute magnitude as a function
of cluster age. The dashed lines are the expected relations for a simple stellar population from  \citet{Girardi2002} with Z=0.004 and masses of 
$10^3$,$10^4$,$10^5$ and $10^6 \Msun$. }
\label{ph_age}
\end{center}
\end{figure}

The reddening-corrected colors of the clusters should correlate with their 
ages and metallicities. The upper panel of Fig. \ref{ph_age} shows the 
variation of integrated 
cluster V-I color with the estimated isochrone age. These have been
reddening corrected using the reddening listed in Table \ref{table}. 
The lines are the expected relations according to simple stellar populations 
\cite{Girardi2002} with Z=0.004 (solid line), Z=0.001 (dotted line) and Z=0.019 
(dashed line). As expected, this plot reveals a positive correlation between
cluster age and V--I color in much the same manner as the models predict.

Our data allow us to compare the luminosities of the clusters with their 
ages in order to derive the cluster masses. The bottom panel of
Fig. \ref{ph_age} shows the integrated absolute magnitudes as a function 
of the estimated isochrone ages. 
These have been corrected using an extinction based on the reddening 
listed in Table \ref{table}. The dashed lines are the expected 
relations for simple stellar populations with Z=0.004 taken from 
\cite{Girardi2002} for masses of $10^3$, $10^4$, $10^5$ and $10^6$ $\Msun$ and assuming a Salpeter initial mass function (IMF). 
This diagram indicates that the clusters
in our sample are consistent with the theoretical predictions of simple stellar 
population fading models. Additionally, the majority of the clusters have 
masses between $5\times10^3$ and $5\times10^4$ \Msun. These estimated 
masses would not change significantly if we were to assume other power-law 
IMFs in constructing the fading lines \citep{Tantalo2005} although we should 
consider them as upper limits in other cases like exponential IMFs 
\citep{Chabrier2001} binary-corrected IMFs \citep{Kroupa1998}. 
It should be noted that the
dearth of clusters older than $\sim$$10^9$ years is attributable to the
fact that our CMDs are generally not deep enough to detect the main sequence
turnoffs of clusters older than this limit.

\section{Summary}
This paper presents \textit{HST/ACS} integrated photometry of 161 star
clusters in M33 as well as individual stellar photometry in twelve fields. 
Forty six clusters of the sample have been previously cataloged by 
\cite{Sarajedini2007}. Color-magnitude diagrams of each cluster have 
been constructed in order to determine ages via isochrone fitting. 
Simple stellar population models reproduce the behavior of the cluster
ages with dereddened V--I color as well as with absolute magnitude.
We have incorporated these new clusters into the 
existing catalog of M33 
clusters established by \citet{Sarajedini2007} increasing  by more than 40\% the 
set of confirmed star clusters in this galaxy. 

\acknowledgments
We thank an anonymous referee for comments that greatly improved this
paper. We are grateful for support from NASA through grant GO-10190 from the Space 
Telescope Science Institute, which is operated by the Association of Universities
for Research in Astronomy, Inc., for NASA under contract NAS5-26555.



\begin{deluxetable}{lcccc}
\tablecaption{Observation Summary}
\tablewidth{0pt}
\tabletypesize{\footnotesize}
\tablehead{
\colhead{Field}	& \colhead{R.A. (J2000.0)}	&
\colhead{Decl. (J2000.0)}	& \colhead{Filter}   &
\colhead{Exp. Time (s)}}
\startdata
D1  & 01 33 49.89 & +30 35 48.12 & F606W &  2 $\times$ 2480, 1 $\times$ 1300, 1 $\times$ 120 \\
    &             &              & F814W &  2 $\times$ 2480, 1 $\times$ 1522, 1 $\times$ 142 \\
    &             &              & F475W &  3 $\times$ 700 \\
D2  & 01 33 39.69 & +30 28 59.98 & F606W &  8 $\times$ 2480, 1 $\times$ 1300, 1 $\times$ 120 \\
    &             &              & F814W & 10 $\times$ 2480, 1 $\times$ 1500, 1 $\times$ 120 \\
    &             &              & F475W &  3 $\times$ 700 \\
D3  & 01 33 20.49 & +30 22 14.99 & F606W &  8 $\times$ 2480, 1 $\times$ 1300, 1 $\times$ 120 \\
    &             &              & F814W & 10 $\times$ 2480, 1 $\times$ 1500, 1 $\times$ 120 \\
    &             &              & F475W &  3 $\times$ 700 \\
D4  & 01 33 07.89 & +30 15 06.98 & F606W &  8 $\times$ 2480, 1 $\times$ 1300, 1 $\times$ 120 \\
    &             &              & F814W & 10 $\times$ 2480, 1 $\times$ 1500, 1 $\times$ 120 \\
    &             &              & F475W &  3 $\times$ 700 \\
F1  & 01 33 40.32 & +30 38 39.89 & F606W &  1 $\times$ 2160 \\
    &             &              & F814W &  1 $\times$ 2160 \\
F2  & 01 33 27.52 & +30 39 15.42 & F606W &  1 $\times$ 2400 \\
    &             &              & F814W &  1 $\times$ 2500 \\   
F3  & 01 34 00.23 & +30 32 56.43 & F606W &  1 $\times$ 2400 \\
    &             &              & F814W &  1 $\times$ 2500 \\
F4  & 01 33 26.04 & +30 30 37.64 & F606W &  1 $\times$ 2160 \\
    &             &              & F814W &  1 $\times$ 2160 \\

F5  & 01 34 11.78 & +30 27 21.94 & F606W &  1 $\times$ 2160 \\
    &             &              & F814W &  1 $\times$ 2160 \\
F6  & 01 33 05.39 & +30 27 11.88 & F606W &  1 $\times$ 2400 \\
    &             &              & F814W &  1 $\times$ 2500 \\   
F7  & 01 33 13.43 & +30 23 29.64 & F606W &  1 $\times$ 2160 \\
    &             &              & F814W &  1 $\times$ 2160 \\
F8  & 01 33 32.92 & +30 17 32.37 & F606W &  1 $\times$ 2400 \\
    &             &              & F814W &  1 $\times$ 2400 

\enddata
\label{observations}
\end{deluxetable}



\begin{deluxetable}{cccccrcrrcc}
\tabletypesize{\footnotesize}
\rotate
\tablewidth{0pt}
\tablecaption{Cluster Properties}
\tablehead{
\colhead{ID}  &\colhead{RA (J2000)} &
\colhead{Dec (J2000)} & \colhead{V}  &
\colhead{(V-I)}   & \colhead{(B-V)}  &
\colhead{E(V-I)}   & \colhead{Log Age\tablenotemark{a}}  &
\colhead{Mass\tablenotemark{b}}  & \colhead{Notes} &
\colhead{Alter. Id \tablenotemark{d}}
}
\startdata
   1  &  1 32 59.38  & 30 26 54.12  &       19.960  &     0.909  &   \nodata  &      0.10  &      8.90  &      6.0  &                   \nodata&  \nodata \\
   2  &  1 32 59.95  & 30 27 19.57  &       19.736  &     1.750  &   \nodata  &      0.10  &      8.55  &      4.0  &                   \nodata       &  \nodata \\
   3  &  1 33  0.37  & 30 26 47.25  &       19.530  &     0.744  &   \nodata  &      0.10  &      8.40  &      4.0  &                  Small       &       31 \\
   4  &  1 33  1.50  & 30 27 47.84  &       19.896  &     0.457  &   \nodata  &      0.06  &      8.20  &      2.0  &                  Small       &  \nodata \\
   5  &  1 33  2.39  & 30 26 55.71  &       19.768  &     0.472  &   \nodata  &      0.10  &      8.30  &      2.5  &                    \nodata      &  \nodata \\
   6  &  1 33  2.98  & 30 26 34.56  &       19.499  &     1.031  &   \nodata  &      0.10  &      8.70  &      6.5  &                    \nodata      &  \nodata \\
   7  &  1 33  3.29  & 30 16  2.26  &       20.348  &     0.816  &     0.403  &      0.10  &      8.75  &      3.0  &                    \nodata      &  \nodata \\
   8  &  1 33  4.35  & 30 27 13.24  &       19.798  &     0.689  &   \nodata  &      0.10  &      8.45  &      3.0  &                    Small      &  \nodata \\
   9  &  1 33  5.62  & 30 14 23.44  &       19.361  &     0.789  &     0.501  &      0.10  &      8.50  &      5.0  &                    \nodata      &  \nodata \\
  10  &  1 33  5.87  & 30 28 49.46  &       19.198  &     0.493  &   \nodata  &      0.10  &      8.25  &      4.0  &                    \nodata      &  \nodata \\
  11  &  1 33  8.11  & 30 27 59.77  &       18.941  &     0.859  &   \nodata  &      0.10  &      8.65  &     10.0  &                     \nodata     & 40 \\
  12  &  1 33  8.93  & 30 16 51.53  &       20.771  &     0.876  &     0.637  &      0.10  &      8.85\tablenotemark{e}  &      2.5  &    \nodata     &  \nodata \\
  13  &  1 33  9.27  & 30 26  6.78  &       19.460  &     0.588  &   \nodata  &      0.10  &      8.40  &      4.0  &                     \nodata     &  \nodata \\
  14  &  1 33  9.77  & 30 22 35.46  &       17.806  &     0.718  &   \nodata  &      0.06  &      8.15  &     10.0  &                     \nodata     &  \nodata \\
  15  &  1 33 10.33  & 30 23  4.26  &       17.530  &    -0.066  &   \nodata  &      0.06  &      7.60\tablenotemark{e}  &      7.5  &    Small     &  \nodata \\
  16  &  1 33 10.38  & 30 15 45.93  &       20.352  &     1.159  &     1.043  &      0.15  &      8.70  &      3.0  &                   Small  &  \nodata \\
  17  &  1 33 12.14  & 30 22 36.97  &       19.818  &     1.181  &   \nodata  &      0.06  &      8.70  &      4.5  &                     \nodata     &  \nodata \\
  18  &  1 33 12.95  & 30 23  7.14  &       18.984  &     0.004  &   \nodata  &      0.06  &      7.70\tablenotemark{e}  &      2.5  &   Small     &  \nodata \\
  19  &  1 33 13.93  & 30 27 59.99  &       19.171  &     0.791  &   \nodata  &   \nodata  &    \nodata  &  \nodata  &                     \nodata    &  \nodata \\
  20  &  1 33 13.96  & 30 14 31.85  &       20.218  &     0.944  &     0.702  &      0.15  &      9.00  &      5.5  &                   \nodata       &  \nodata \\
  21  &  1 33 14.04  & 30 15 16.47  &       20.075  &     0.770  &     0.511  &      0.20  &      8.65  &      4.0  &                   Small  &  \nodata \\
  22  &  1 33 14.33  & 30 28 22.42  &       18.252  &     1.117  &   \nodata  &      0.06  &      8.80  &     25.0  &                    \nodata       &	49 \\
  23  &  1 33 14.71  & 30 23 19.05  &       19.328  &     0.908  &   \nodata  &      0.15  &      8.80  &      9.0  &                    \nodata       &  \nodata \\
  24  &  1 33 15.22  & 30 21 14.07  &       19.355  &     0.148  &    -0.092  &      0.06  &      8.20  &      3.0  &                     \nodata      &  \nodata \\
  25  &  1 33 16.10  & 30 20 56.55  &       18.245  &     0.566  &     0.244  &      0.10  &      8.30  &     10.0  &                     \nodata      &	53 \\
  26  &  1 33 19.09  & 30 30 10.68  &       18.917  &     0.833  &   \nodata  &      0.06  &      8.50  &      7.5  &                     \nodata      &  \nodata \\
  27  &  1 33 19.16  & 30 23 22.62  &       18.470  &     0.886  &   \nodata  &      0.06  &      9.00  &     25.0  &                      \nodata     &	57 \\
  28  &  1 33 19.91  & 30 30 20.22  &       19.359  &     1.180  &   \nodata  &      0.15  &      8.80  &      9.0  &                     \nodata      &  \nodata \\
  29  &  1 33 21.09  & 30 37 55.77  &       18.943  &     0.804  &   \nodata  &      0.10  &      8.55  &      8.5  &                     \nodata      &	61 \\
  30  &  1 33 21.31  & 30 20 31.88  &       18.184  &     0.429  &     0.236  &      0.06  &      8.20  &      8.5  &                     \nodata      &  \nodata \\
  31  &  1 33 21.34  & 30 31  0.89  &       19.351  &     1.091  &   \nodata  &      0.15  &      8.80  &      9.0  &                      \nodata     &  \nodata \\
  32  &  1 33 21.35  & 30 31 31.31  &       18.906  &     0.871  &   \nodata  &   \nodata  &    \nodata  &  \nodata  &                    \nodata       &  \nodata \\
  33  &  1 33 21.38  & 30 31 12.56  &       18.585  &     0.551  &   \nodata  &      0.06  &      8.20  &      6.0  &                      \nodata     &  \nodata \\
  34  &  1 33 21.57  & 30 31 50.96  &       18.636  &     1.276  &   \nodata  &   \nodata  &    \nodata  &  \nodata  &                    \nodata       &	 62 \\
  35  &  1 33 21.59  & 30 37 48.69  &       19.027  &     0.675  &   \nodata  &      0.15  &      8.40  &      6.5  &                     \nodata      &	63 \\
  36  &  1 33 22.09  & 30 40 26.20  &       18.339  &     0.382  &   \nodata  &   \nodata  &    \nodata  &  \nodata  &                   Small       &	 66 \\
  37  &  1 33 22.14  & 30 38 28.06  &       18.999  &     0.932  &   \nodata  &      0.06  &      8.50  &      7.0  &                      Small       &  \nodata \\
  38  &  1 33 22.37  & 30 30 14.12  &       17.426  &     0.484  &   \nodata  &      0.06  &      8.00  &     15.0  &                     \nodata      &	69 \\
  39  &  1 33 22.67  & 30 38  0.06  &       19.235  &     1.049  &   \nodata  &      0.06  &      9.20  &     20.0  &                      Small       &  \nodata \\
  40  &  1 33 22.76  & 30 38 19.87  &       19.050  &     1.050  &   \nodata  &      0.10  &      8.35  &      5.5  &                    Small  &  \nodata \\
  41  &  1 33 23.10  & 30 32 22.69  &       18.946  &     0.909  &   \nodata  &      0.15  &      8.35  &      6.5  &                     \nodata      &       71 \\
  42  &  1 33 23.83  & 30 40 26.26  &       19.004  &     0.753  &   \nodata  &      0.06  &      8.50  &      7.0  &                    Small  &       74 \\
  43  &  1 33 23.84  & 30 39 36.58  &       18.882  &     0.636  &   \nodata  &      0.15  &      8.30  &      6.0  &                   \nodata        &  \nodata \\
  44  &  1 33 24.61  & 30 37 49.88  &       18.605  &     0.784  &   \nodata  &      0.10  &      8.45  &      9.5  &                    \nodata       &  \nodata \\
  45  &  1 33 25.16  & 30 32 18.06  &       20.064  &     1.031  &   \nodata  &      0.10  &      8.70  &      4.0  &                    \nodata       &  \nodata \\
  46  &  1 33 25.33  & 30 23 38.97  &       20.091  &     0.744  &     0.446  &      0.10  &      8.45  &      2.5  &                   Small   &  \nodata \\
  47  &  1 33 25.73  & 30 18  1.12  &       19.719  &     0.157  &   \nodata  &      0.06  &      8.25  &      2.5  &                   \nodata        &  \nodata \\
  48  &  1 33 25.76  & 30 31 19.80  &       19.610  &     0.927  &   \nodata  &      0.10  &      8.60  &      5.0  &                   \nodata        &  \nodata \\
  49  &  1 33 26.50  & 30 30  1.71  &       19.448  &     0.909  &   \nodata  &      0.10  &      8.45  &      4.5  &                   Small   &  \nodata \\
  50  &  1 33 26.52  & 30 37 55.19  &       19.530  &     0.612  &   \nodata  &      0.06  &      8.45  &      4.0  &                   Small   &  \nodata \\
  51  &  1 33 26.63  & 30 31 30.09  &       18.114  &     0.546  &   \nodata  &      0.10  &      8.15  &      9.0  &                    \nodata       &  \nodata \\
  52  &  1 33 27.43  & 30 30  3.38  &       18.574  &     0.900  &   \nodata  &      0.15  &      8.50  &     12.0  &                    \nodata       &  \nodata \\
  53  &  1 33 27.86  & 30 38 49.00  &       18.004  &     0.773  &   \nodata  &   \nodata  &    \nodata  &  \nodata  &                   \nodata        &  \nodata \\
  54  &  1 33 29.26  & 30 29 13.67  &       18.918  &     0.681  &   \nodata  &      0.06  &      8.25  &      5.0  &                     Small  &  \nodata \\
  55  &  1 33 29.47  & 30 30  1.88  &       17.932  &     0.544  &   \nodata  &      0.10  &      8.00  &      9.5  &                    \nodata       &	98 \\
  56  &  1 33 29.55  & 30 29 34.38  &       19.206  &     0.796  &   \nodata  &      0.06  &      8.25  &      3.5  &                    \nodata       &  \nodata \\
  57  &  1 33 29.87  & 30 31 16.14  &       18.944  &     0.529  &   \nodata  &      0.06  &      8.35  &      5.5  &                    Small       &  \nodata \\
  58  &  1 33 30.18  & 30 29 33.65  &       19.514  &     0.798  &   \nodata  &      0.06  &      8.25  &      3.0  &                    \nodata       &  \nodata \\
  59  &  1 33 30.34  & 30 37 43.37  &       18.834  &     0.771  &   \nodata  &      0.10  &      8.40  &      7.0  &                    \nodata       &  \nodata \\
  60  &  1 33 30.83  & 30 29 14.10  &       18.218  &     0.668  &   \nodata  &      0.10  &      7.65  &      4.5  &                   Small       &  \nodata \\
  61  &  1 33 31.31  & 30 40 20.54  &       17.966  &     0.844  &   \nodata  &      0.20  &      8.45  &     20.0  &                    \nodata       &      110 \\
  62  &  1 33 31.44  & 30 28 42.00  &       18.398  &     0.896  &     0.119  &      0.20  &      8.10  &      7.0  &                     \nodata      &  \nodata \\
  63  &  1 33 32.09  & 30 40 32.02  &       18.494  &     0.344  &   \nodata  &      0.10  &      7.80  &      4.5  &                     \nodata      &      114 \\
  64  &  1 33 32.18  & 30 30 17.21  &       18.158  &     0.459  &   \nodata  &      0.06  &      8.15  &      8.0  &                     \nodata      &  \nodata \\
  65  &  1 33 32.35  & 30 38 24.70  &       18.376  &     0.828  &   \nodata  &      0.20  &      8.30  &     10.0  &                     \nodata      &      117 \\
  66  &  1 33 32.51  & 30 39 24.66  &       18.835  &     1.028  &   \nodata  &      0.10  &      8.45  &      8.0  &                     \nodata      &      118 \\
  67  &  1 33 32.75  & 30 31 44.91  &       19.395  &     1.221  &   \nodata  &      0.15  &      8.65  &      7.0  &                        Small  &      120 \\
  68  &  1 33 32.87  & 30 39 31.82  &       18.940  &     0.626  &   \nodata  &      0.10  &      8.40  &      6.5  &                      \nodata       &  \nodata \\
  69  &  1 33 32.88  & 30 15 46.76  &       19.893  &     0.287  &   \nodata  &      0.10  &      8.15  &      2.0  &                      \nodata       &  \nodata \\
  70  &  1 33 33.37  & 30 29 45.73  &       19.027  &     0.938  &    -0.107  &      0.06  &      7.85  &      3.0  &                      Small      &  \nodata \\
  71  &  1 33 33.63  & 30 28  9.53  &       18.167  &     0.326  &    -0.047  &      0.10  &      8.10  &      8.0  &                      \nodata       &  \nodata \\
  72  &  1 33 33.64  & 30 40  3.15  &       18.279  &     0.747  &   \nodata  &      0.10  &      8.35  &     10.0  &                      \nodata       &      124 \\
  73  &  1 33 33.88  & 30 39 54.12  &       18.524  &     0.665  &   \nodata  &      0.06  &      8.10  &      5.5  &                         Small  &  \nodata \\
  74  &  1 33 35.54  & 30 38 36.90  &       17.190  &     0.261  &   \nodata  &   \nodata  &    \nodata  &  \nodata  &                     \nodata      &	132 \\
  75  &  1 33 35.86  & 30 27 44.54  &       19.428  &     0.622  &     0.307  &      0.06  &      8.40  &      4.0  &                      \nodata     &  \nodata \\
  76  &  1 33 36.24  & 30 27 56.79  &       19.212  &     0.960  &     0.472  &      0.10  &      8.60  &      7.0  &                      \nodata     &  \nodata \\
  77  &  1 33 36.38  & 30 15 32.61  &       20.602  &     0.904  &   \nodata  &      0.10  &      8.70  &      2.5  &                      \nodata     &  \nodata \\
  78  &  1 33 36.66  & 30 27  8.04  &       18.468  &     0.970  &     0.278  &      0.15  &      8.45  &     12.0  &                      \nodata     &      136 \\
  79  &  1 33 37.33  & 30 38 38.67  &       18.805  &     0.815  &   \nodata  &      0.10  &      8.20  &      5.0  &                          Small  &  \nodata \\
  80  &  1 33 37.51  & 30 28  4.58  &       17.685  &     0.809  &     0.250  &      0.10  &      8.25  &     15.0  &                     \nodata       &	141 \\
  81  &  1 33 37.90  & 30 38  2.42  &       17.465  &     1.251  &   \nodata  &      0.10  &      8.60  &     35.0  &                     \nodata       &	143 \\
  82  &  1 33 37.94  & 30 18 51.74  &       19.889  &     0.646  &   \nodata  &      0.10  &      8.65  &      4.0  &                     \nodata       &  \nodata \\
  83  &  1 33 38.29  & 30 17 35.99  &       19.927  &     1.482  &   \nodata  &      0.06  &      8.50  &      3.0  &                     \nodata       &  \nodata \\
  84  &  1 33 39.08  & 30 38 25.21  &       17.501  &     1.322  &   \nodata  &      0.06  &      7.70  &      9.0  &                     \nodata       &  \nodata \\
  85  &  1 33 39.47  & 30 28  7.29  &       19.405  &     1.361  &     0.761  &      0.10  &      8.90  &      9.5  &                          Small  &  \nodata \\
  86  &  1 33 39.63  & 30 31  9.16  &       16.575  &     0.331  &    -0.162  &      0.06  &      7.90  &     30.0  &                     \nodata       &	152 \\
  87  &  1 33 39.84  & 30 38 26.43  &       16.013  &     0.879  &   \nodata  &      0.06  &      7.25  &     20.0  &                     \nodata       &	154 \\
  88  &  1 33 40.12  & 30 37 45.82  &       18.035  &     0.793  &   \nodata  &      0.10  &      8.10  &      9.0  &                     \nodata       &  \nodata \\
  89  &  1 33 41.13  & 30 29 53.76  &       18.961  &     0.854  &     0.458  &      0.15  &      8.35  &      6.0  &                     \nodata       &	159 \\
  90  &  1 33 41.43  & 30 31 13.39  &      \nodata  &     0.247  &    -0.136  &      0.10  &      7.85  &  \nodata  &                           Edge  &  \nodata \\
  91  &  1 33 41.49  & 30 30 24.11  &       18.975  &     0.621  &     0.130  &      0.06  &      8.15  &      4.0  &                          Small  &      162 \\
  92  &  1 33 41.53  & 30 28  9.24  &       19.005  &     0.277  &    -0.046  &      0.06  &      8.25  &      4.5  &                     \nodata       &	163 \\
  93  &  1 33 41.69  & 30 34 57.49  &       19.446  &     0.636  &    -0.267  &      0.10  &      8.05  &      2.5  &                     Small         &  \nodata \\
  94  &  1 33 41.77  & 30 29 32.49  &       18.146  &     0.600  &     0.161  &      0.10  &      8.05  &      8.0  &                     \nodata       &  \nodata \\
  95  &  1 33 41.98  & 30 38 20.55  &       18.342  &     0.933  &   \nodata  &      0.15  &      8.35  &     10.0  &                     \nodata       &  \nodata \\
  96  &  1 33 42.62  & 30 34 58.93  &       19.579  &     1.350  &     0.324  &      0.06  &      8.85  &      7.0  &                     \nodata       &  \nodata \\
  97  &  1 33 42.62  & 30 38 22.11  &       18.350  &     0.747  &   \nodata  &      0.10  &      8.05  &      7.0  &                        Small  &  \nodata \\
  98  &  1 33 42.89  & 30 27 46.46  &       19.226  &     1.155  &     0.334  &      0.10  &      8.60  &      7.0  &                        Small  &  \nodata \\
  99  &  1 33 42.90  & 30 38 30.91  &       18.313  &     0.889  &   \nodata  &   \nodata  &    \nodata  &  \nodata  &                       Small  &  \nodata \\
 100  &  1 33 43.47  & 30 28  8.20  &       18.996  &     1.104  &     0.600  &      0.06  &      8.25  &      4.5  &                      Small       &  \nodata \\
 101  &  1 33 43.53  & 30 27 58.28  &       18.118  &     0.693  &    -0.086  &      0.10  &      8.15  &      9.0  &                     \nodata       &  \nodata \\
 102  &  1 33 43.93  & 30 36 13.09  &       19.049  &     1.029  &    -0.127  &      0.06  &      8.45  &      6.0  &                      \nodata      &  \nodata \\
 103  &  1 33 44.00  & 30 30  0.79  &       18.164  &     0.669  &     0.169  &      0.10  &      8.20  &      9.0  &                     \nodata       &	174 \\
 104  &  1 33 44.15  & 30 35 25.44  &       19.062  &     0.928  &    -0.032  &      0.20  &      8.30  &      5.5  &                     \nodata       &  \nodata \\
 105  &  1 33 44.35  & 30 38  5.36  &       18.194  &     0.376  &   \nodata  &      0.06  &      8.20  &      8.5  &                     \nodata       &  \nodata \\
 106  &  1 33 44.42  & 30 37 52.96  &       17.314  &     0.681  &   \nodata  &      0.15  &      8.10  &     20.0  &                     \nodata       &	176 \\
 107  &  1 33 44.49  & 30 37  6.90  &       19.723  &     0.909  &    -0.102  &      0.06  &      8.50  &      3.5  &                      \nodata      &  \nodata \\
 108  &  1 33 44.49  & 30 36 43.50  &       19.463  &     1.298  &    -0.033  &      0.10  &      8.20  &      3.0  &                         Small     &  \nodata \\
 109  &  1 33 44.52  & 30 39 19.62  &       18.349  &     1.053  &   \nodata  &      0.15  &      8.30  &     10.0  &                      \nodata      &  \nodata \\
 110  &  1 33 44.56  & 30 37 34.16  &       18.613  &     0.817  &   \nodata  &      0.15  &      8.15  &      6.0  &                       \nodata     &  \nodata \\
 111  &  1 33 44.64  & 30 36 35.65  &       18.303  &     0.558  &    -0.799  &      0.06  &      7.50\tablenotemark{e}  &      3.0  &       Small      &  \nodata \\
 112  &  1 33 44.84  & 30 34 38.87  &       18.808  &     0.564  &    -0.333  &      0.06  &      7.70  &      2.5  &                       Small      &  \nodata \\
 113  &  1 33 45.96  & 30 36 49.35  &       18.561  &     0.562  &    -0.397  &      0.10  &      8.20  &      6.5  &                      \nodata      &  \nodata \\
 114  &  1 33 46.18  & 30 35 27.75  &       18.323  &     0.764  &    -0.287  &      0.06  &      7.85  &      5.5  &                       Small      &  \nodata \\
 115  &  1 33 46.75  & 30 35 59.00  &       19.829  &     1.218  &    -0.057  &      0.15  &      8.35  &      3.0  &                      Small     &  \nodata \\
 116  &  1 33 47.35  & 30 39 59.41  &       18.405  &     0.967  &   \nodata  &      0.10  &      8.45  &     12.0  &                     \nodata       &  \nodata \\
 117  &  1 33 47.44  & 30 39  3.71  &       17.975  &     0.740  &   \nodata  &      0.06  &      8.05  &      9.0  &                     \nodata       &  \nodata \\
 118  &  1 33 48.05  & 30 39 29.01  &       18.151  &     1.196  &   \nodata  &   \nodata  &    \nodata  &  \nodata  &                     \nodata       &  \nodata \\
 119  &  1 33 48.55  & 30 37  5.08  &       18.547  &     1.121  &    -0.318  &      0.25  &      8.00  &      6.5  &                     \nodata       &  \nodata \\
 120  &  1 33 48.58  & 30 35 23.25  &       19.093  &     0.901  &    -0.156  &      0.15  &      8.50  &      7.0  &                     \nodata       &  \nodata \\
 121  &  1 33 49.53  & 30 28 42.42  &       19.198  &     0.630  &     0.312  &      0.20  &      8.55  &      7.5  &                     \nodata       &  \nodata \\
 122  &  1 33 49.58  & 30 34 25.55  &       18.361  &     0.960  &    -0.213  &      0.20  &      8.15  &      8.0  &                     \nodata       &	191 \\
 123  &  1 33 49.87  & 30 36 34.28  &       18.022  &     0.949  &    -0.346  &      0.06  &      8.00  &      8.5  &                      \nodata      &  \nodata \\
 124  &  1 33 50.15  & 30 34 18.67  &       19.098  &     0.922  &    -0.026  &      0.20  &      8.45  &      7.0  &                      \nodata      &	192 \\
 125  &  1 33 50.98  & 30 35 24.43  &       19.688  &     1.042  &    -0.026  &   \nodata  &    \nodata  &  \nodata  &                  Small  &  \nodata \\
 126  &  1 33 51.20  & 30 34 12.97  &       18.479  &     0.747  &    -0.234  &      0.10  &      8.10  &      6.0  &                  \nodata        &      201 \\
 127  &  1 33 51.78  & 30 34 18.44  &       19.630  &     0.959  &     0.023  &   \nodata  &    \nodata  &  \nodata  &                  Small  &  \nodata \\
 128  &  1 33 52.34  & 30 35  0.56  &       18.937  &     0.743  &    -0.186  &      0.20  &      8.10  &      4.5  &                  \nodata         &      207 \\
 129  &  1 33 52.35  & 30 34 20.92  &       18.967  &     0.866  &    -0.124  &      0.20  &      8.35  &      6.5  &                  \nodata         &      208 \\
 130  &  1 33 53.30  & 30 33  2.89  &       18.938  &     1.007  &   \nodata  &      0.15  &      8.40  &      7.0  &                  \nodata         &      212 \\
 131  &  1 33 54.16  & 30 36  6.73  &       19.437  &     0.849  &    -0.083  &      0.20  &      8.30  &      4.0  &                  \nodata         &  \nodata \\
 132  &  1 33 54.39  & 30 32 23.90  &       18.611  &     0.997  &   \nodata  &      0.15  &      8.60  &     14.0  &                   \nodata        &  \nodata \\
 133  &  1 33 54.59  & 30 34 48.08  &       19.167  &     1.237  &    -0.153  &      0.20  &      8.40  &      6.0  &                  \nodata         &      217 \\
 134  &  1 33 54.66  & 30 32 15.71  &       17.644  &     0.392  &   \nodata  &      0.06  &      8.05  &     12.0  &                  \nodata         &      220 \\
 135  &  1 33 54.91  & 30 32 14.46  &       18.454  &     0.596  &   \nodata  &      0.10  &      8.30  &      8.5  &                   \nodata        &      221 \\
 136  &  1 33 55.64  & 30 33 44.62  &       18.553  &     0.934  &   \nodata  &      0.10  &      8.35  &      8.5  &                    Small        &  \nodata \\
 137  &  1 33 56.42  & 30 36 10.42  &       18.702  &     1.012  &    -0.167  &      0.20  &      8.10  &      5.5  &                   \nodata        &      231 \\
 138  &  1 33 57.08  & 30 34 15.09  &       18.286  &     0.375  &   \nodata  &      0.06  &      7.85  &      5.5  &                    Small        &  \nodata \\
 139  &  1 33 57.75  & 30 33 25.67  &       17.033  &     0.406  &   \nodata  &      0.10  &      7.90  &     20.0  &                    \nodata       &      243 \\
 140  &  1 33 57.80  & 30 35 31.57  &       18.221  &     1.072  &     0.072  &      0.15  &      8.45  &     15.0  &                   \nodata        &      241 \\
 141  &  1 33 57.84  & 30 32 21.86  &       18.749  &     1.940  &   \nodata  &      0.15  &      8.60  &     12.0  &                   \nodata        &  \nodata \\
 142  &  1 33 58.87  & 30 33 29.30  &       18.948  &     0.640  &   \nodata  &      0.10  &      8.35  &      6.0  &                   Small        &  \nodata \\
 143  &  1 33 59.19  & 30 33 45.56  &       18.126  &     0.483  &   \nodata  &      0.15  &      8.25  &     12.0  &                   Small       &  \nodata \\
 144  &  1 33 59.53  & 30 36 24.30  &       18.555  &     0.581  &    -0.304  &      0.10  &      8.10  &      5.5  &                   \nodata        &  \nodata \\
 145  &  1 33 59.70  & 30 32  0.31  &       19.271  &     1.333  &   \nodata  &   \nodata  &    \nodata  &  \nodata  &                  Small         &  \nodata \\
 146  &  1 33 59.88  & 30 33 54.27  &       16.273  &     0.960  &   \nodata  &      0.06  &      7.50  &     20.0  &                   \nodata        &      260 \\
 147  &  1 34  1.63  & 30 32 25.60  &       18.493  &     0.948  &   \nodata  &      0.10  &      8.50  &     12.0  &                   \nodata        &      271 \\
 148  &  1 34  1.74  & 30 34  6.17  &       18.995  &     1.020  &   \nodata  &      0.12  &      8.90  &     15.0  &                        Small  &  \nodata \\
 149  &  1 34  2.25  & 30 32 37.75  &       18.659  &     1.187  &   \nodata  &   \nodata  &    \nodata  &  \nodata  &                   \nodata        &  \nodata \\
 150  &  1 34  3.25  & 30 27 56.15  &       17.842  &     0.500  &   \nodata  &      0.06  &      8.10  &     10.0  &                    \nodata       &  \nodata \\
 151  &  1 34  3.44  & 30 33 41.47  &       18.943  &     0.511  &   \nodata  &      0.10  &      8.30  &      5.5  &                    \nodata       &  \nodata \\
 152  &  1 34  4.60  & 30 27 21.66  &       19.722  &     0.814  &   \nodata  &      0.06  &      8.80  &      6.0  &                    \nodata       &  \nodata \\
 153  &  1 34  6.68  & 30 28  4.43  &       19.187  &     0.674  &   \nodata  &      0.06  &      8.20  &      3.5  &                    \nodata       &  \nodata \\
 154  &  1 34  6.85  & 30 32  0.08  &       18.314  &     0.798  &   \nodata  &      0.10  &      8.55  &     15.0  &                    \nodata       &      306 \\
 155  &  1 34  7.96  & 30 31 18.36  &      \nodata  &   \nodata  &   \nodata  &   \nodata  &    \nodata  &  \nodata  &                        Edge  &  \nodata \\
 156  &  1 34  9.64  & 30 25  5.03  &       19.626  &     0.335  &   \nodata  &      0.06  &      8.50  &      4.0  &                     \nodata        &  \nodata \\
 157  &  1 34 12.88  & 30 28 46.31  &       18.037  &     0.619  &   \nodata  &      0.10  &      7.80  &     6.5  &                     \nodata        &  \nodata \\
 158  &  1 34 13.48  & 30 28 44.11  &       19.721  &     0.873  &   \nodata  &      0.10  &      8.75  &      5.5  &                     \nodata        &  \nodata \\
 159  &  1 34 13.93  & 30 27 59.02  &       18.342  &     1.767  &   \nodata  &      0.06  &      7.95\tablenotemark{e}  &      6.0  &    \nodata        &      346 \\
 160  &  1 34 17.72  & 30 27  8.99  &       20.037  &     0.657  &   \nodata  &      0.15  &      8.40  &      2.5  &                     \nodata        &  \nodata \\
 161  &  1 34 18.62  & 30 27 19.31  &       19.729  &     0.945  &   \nodata  &      0.10  &      8.85  &      6.5  &                      \nodata       &  \nodata \\
\enddata
\tablecomments{Units of RA are hours, minutes, and seconds, and
units of Dec are degrees, arcminutes, and arcseconds.}
\tablenotetext{a}{Units of age are in years.}
\tablenotetext{b}{Units of mass are $\times10^{3} \Msun$}
\tablenotetext{c}{Error in Log Age of $\pm$0.1.}
\tablenotetext{d}{Identification Number in \cite{Sarajedini2007}}
\label{table}
\end{deluxetable}


\begin{deluxetable}{lccccc}
\tablecaption{Cross identification with \cite{Zlocz2008} }
\tablewidth{0pt}
\tabletypesize{\footnotesize}
\tablehead{
\colhead{ID(ZKH)} & \colhead{ID(us)}	& 
 \colhead{R.A. (J2000.0)} &\colhead{Decl. (J2000.0)}& 
\colhead{Notes}   & \colhead{Type}\tablenotemark{a}}
\startdata
25-1-009&        1&       1      32      59.38&      30      26      54.12& \nodata&	    0\\
25-1-008&        2&       1      32      59.95&      30      27      19.57& \nodata&	    0\\
25-1-003&        5&       1      33      02.39&      30      26      55.71& \nodata&	    0\\
25-1-001&        6&       1      33      02.98&      30      26      34.56& \nodata&	    0\\
34-2-001&        7&       1      33      03.29&      30      16      02.26& \nodata&	   1\\
33-4-018&        9&       1      33      05.62&      30      14      23.44& \nodata&	    1\\
33-5-022&       12&       1      33      08.93&      30      16      51.53& \nodata&	    0\\
33-6-016&       14&       1      33      09.77&      30      22      35.46& \nodata&	    1\\
33-5-019&       16&       1      33      10.38&      30      15      45.93& \nodata&	    0\\
33-6-014&       17&       1      33      12.14&      30      22      36.97& \nodata &	    1\\
33-6-010&       18&       1      33      12.95&      30      23      07.14& Small&  0\\
33-5-014&       20&       1      33      13.96&      30      14      31.85& \nodata&       0\\
33-5-013&       21&       1      33      14.04&      30      15      16.47& Small&0\\
33-6-009&       23&       1      33      14.71&      30      23      19.05& \nodata&	   1\\
33-6-008&       24&       1      33      15.22&      30      21      14.07& \nodata&	   0\\
33-6-006&       25&       1      33      16.10&      30      20      56.55& \nodata&	   2\\
33-3-021&       27&       1      33      19.16&      30      23      22.62& \nodata&	   1\\
33-3-020&       30&       1      33      21.31&      30      20      31.88& \nodata&	   0\\
33-2-010&       47&       1      33      25.73&      30      18      01.12& \nodata&	  0\\
33-2-003&       69&       1      33      32.88&      30      15      46.76& \nodata&	   0\\
32-5-024&       77&       1      33      36.38&      30      15      32.61& \nodata&	   0\\
\enddata
\tablecomments{Units of RA are hours, minutes, and seconds, and
units of Dec are degrees, arcminutes, and arcseconds.}
\tablenotetext{a}{Proposed classification in ZKH: -1 galaxy, 0 unclassified, 1 likely stellar cluster and  2 an already known high confidence cluster included in \cite{Sarajedini2007}}
\label{cross_id}
\end{deluxetable}



\begin{figure}
\begin{center}
\epsscale{0.9}
\includegraphics[bb= 59 409 622 1038,width=0.41\textwidth,clip=true]{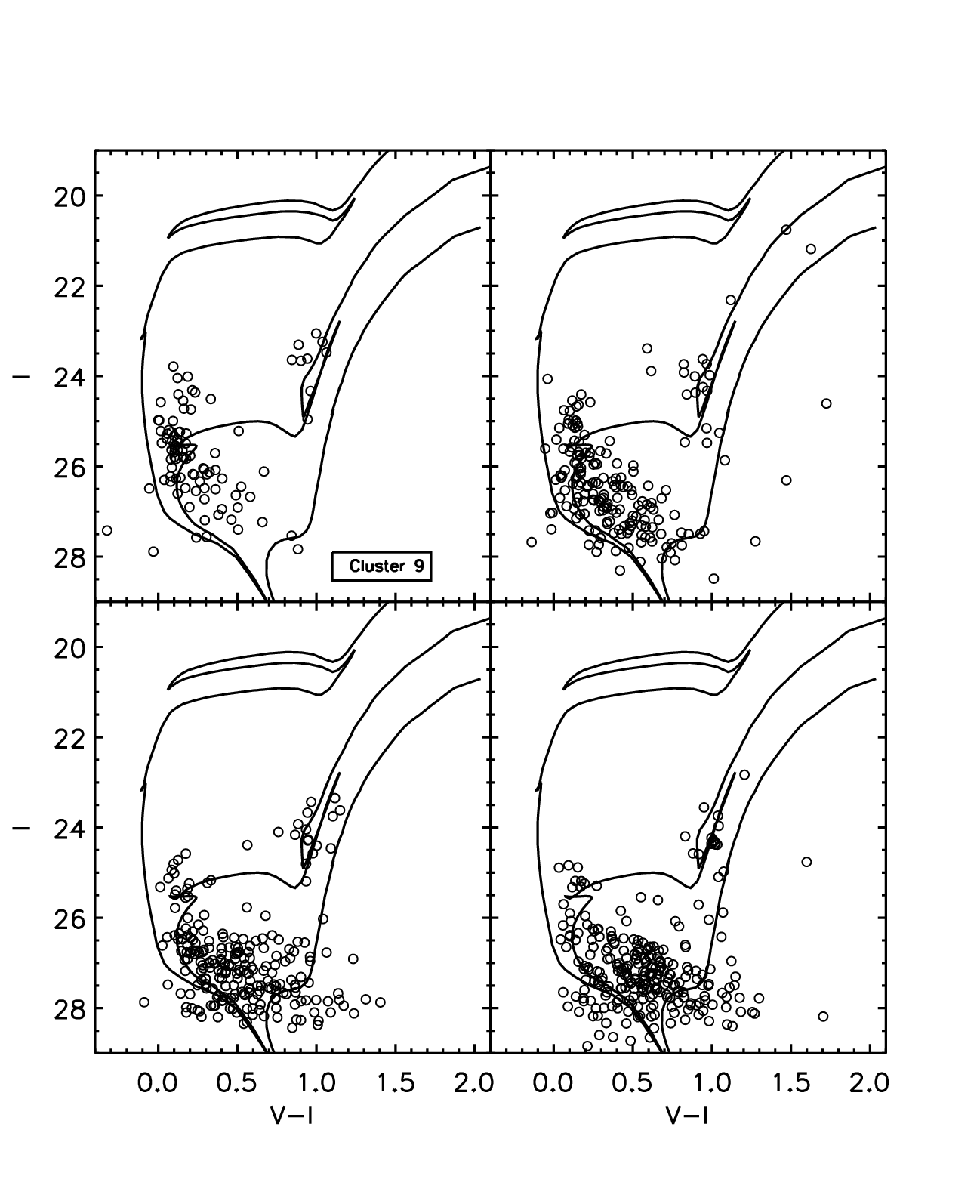}
\hfil
\includegraphics[bb=87 377 660 1110,width=0.41\textwidth,clip=true]{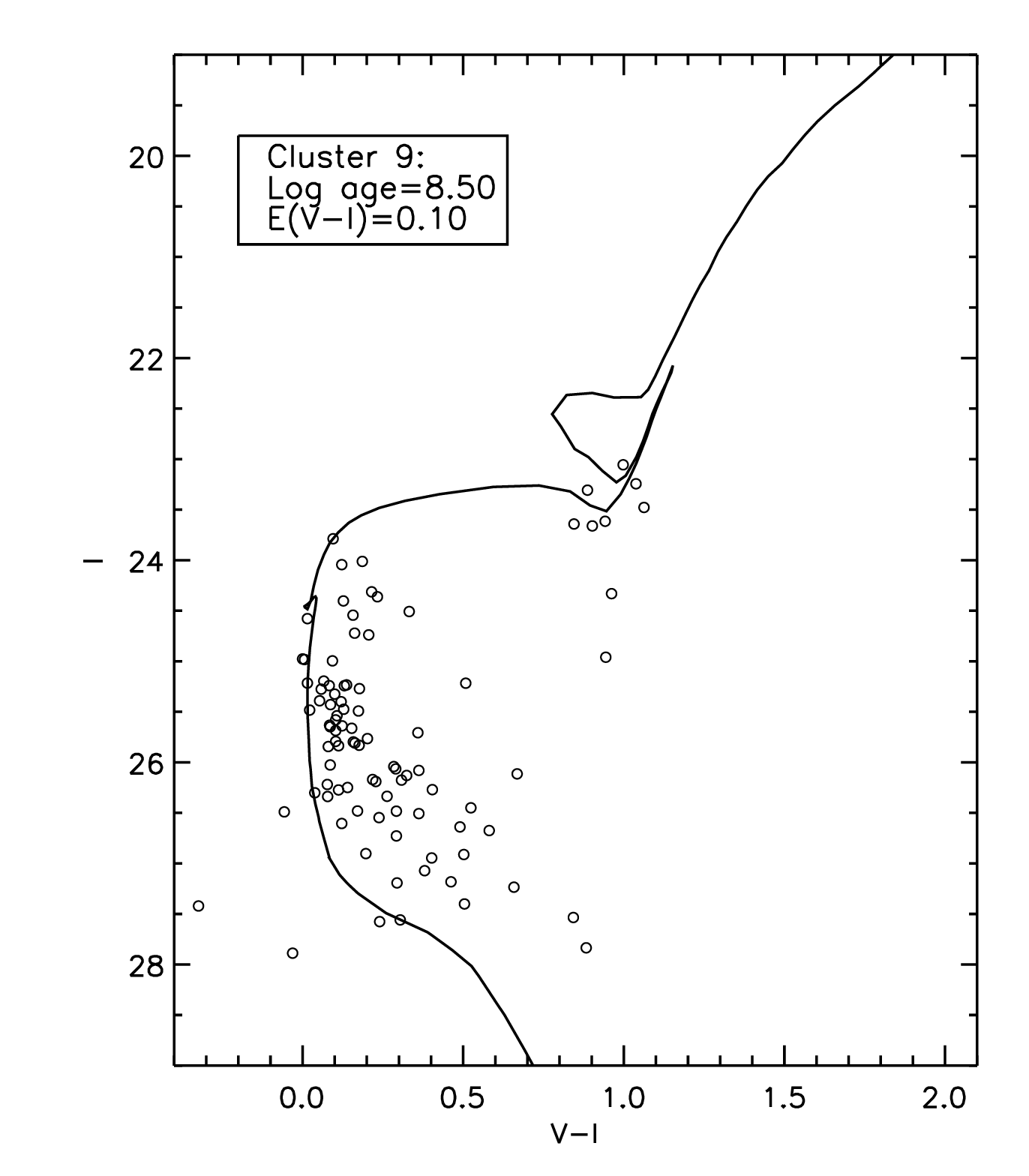}
\caption{Same as Fig. \ref{cmd1}, but for cluster 9.}
\label{cmd2}
\end{center}
\end{figure}

\begin{figure}
\begin{center}
\epsscale{0.9}
\includegraphics[bb= 59 409 622 1038,width=0.41\textwidth,clip=true]{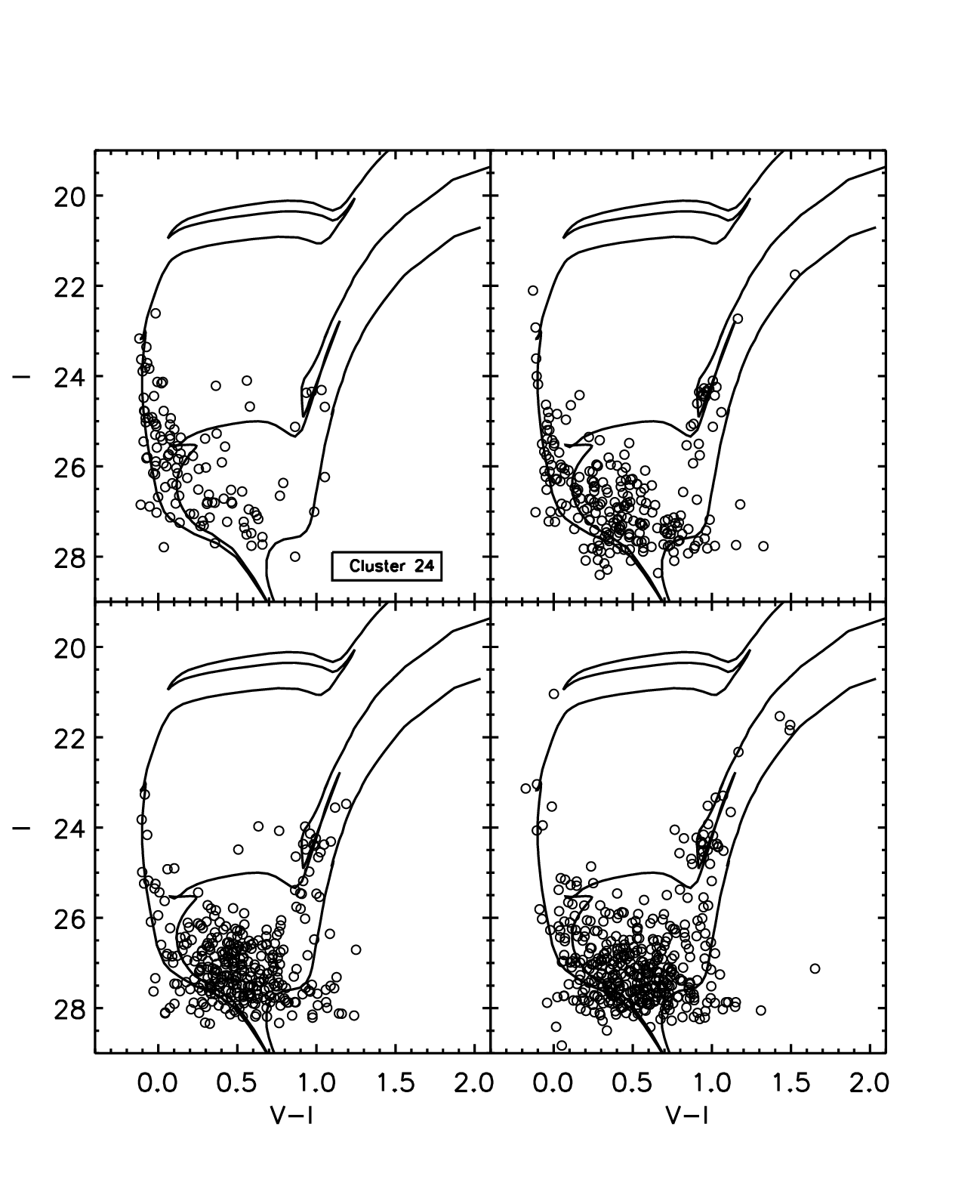}
\hfil
\includegraphics[bb=87 377 660 1110,width=0.41\textwidth,clip=true]{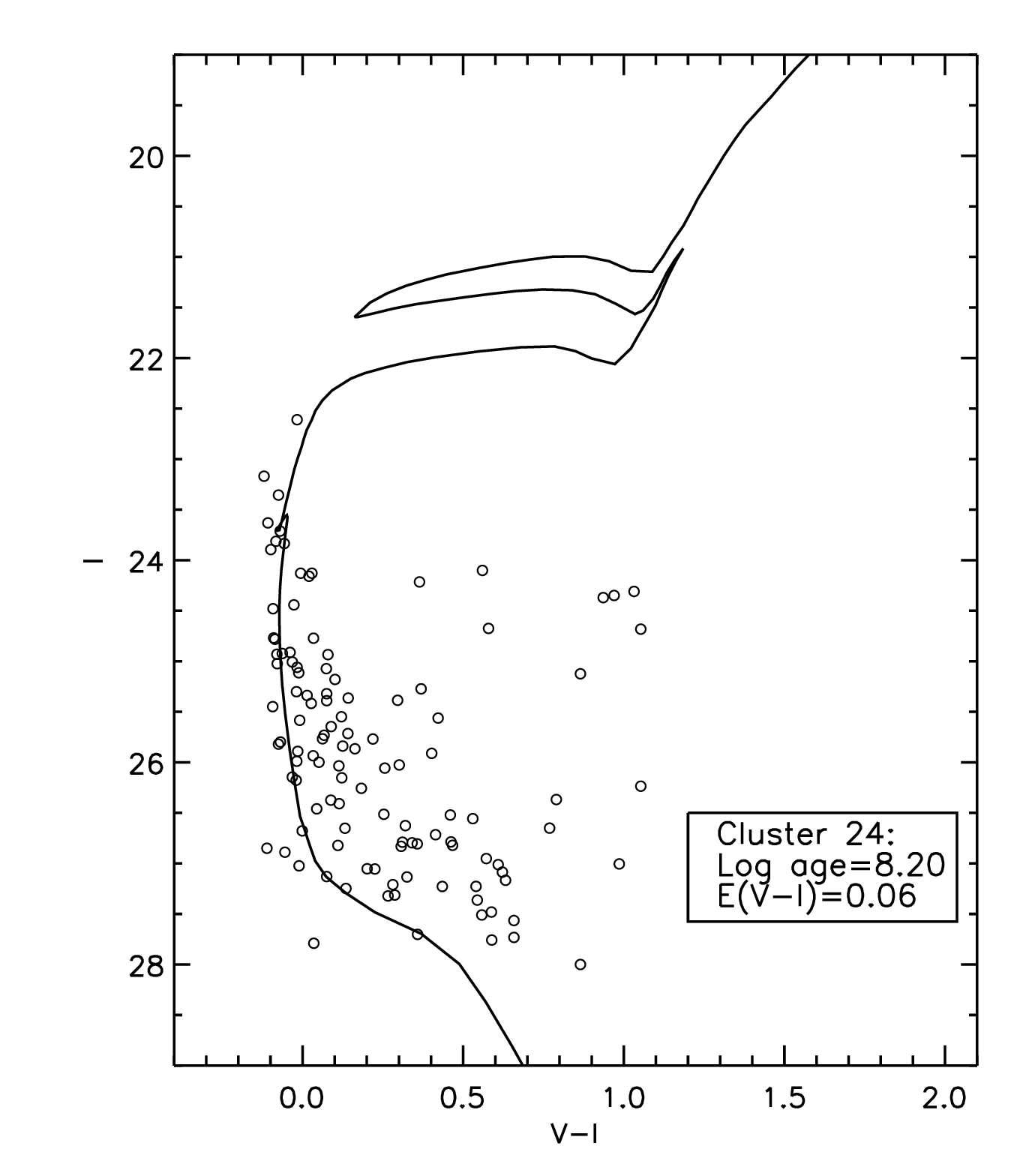}
\caption{Same as Fig. \ref{cmd1}, but for cluster 24.}
\label{cmd3}
\end{center}
\end{figure}

\begin{figure}
\begin{center}
\epsscale{0.9}
\includegraphics[bb= 59 409 622 1038,width=0.41\textwidth,clip=true]{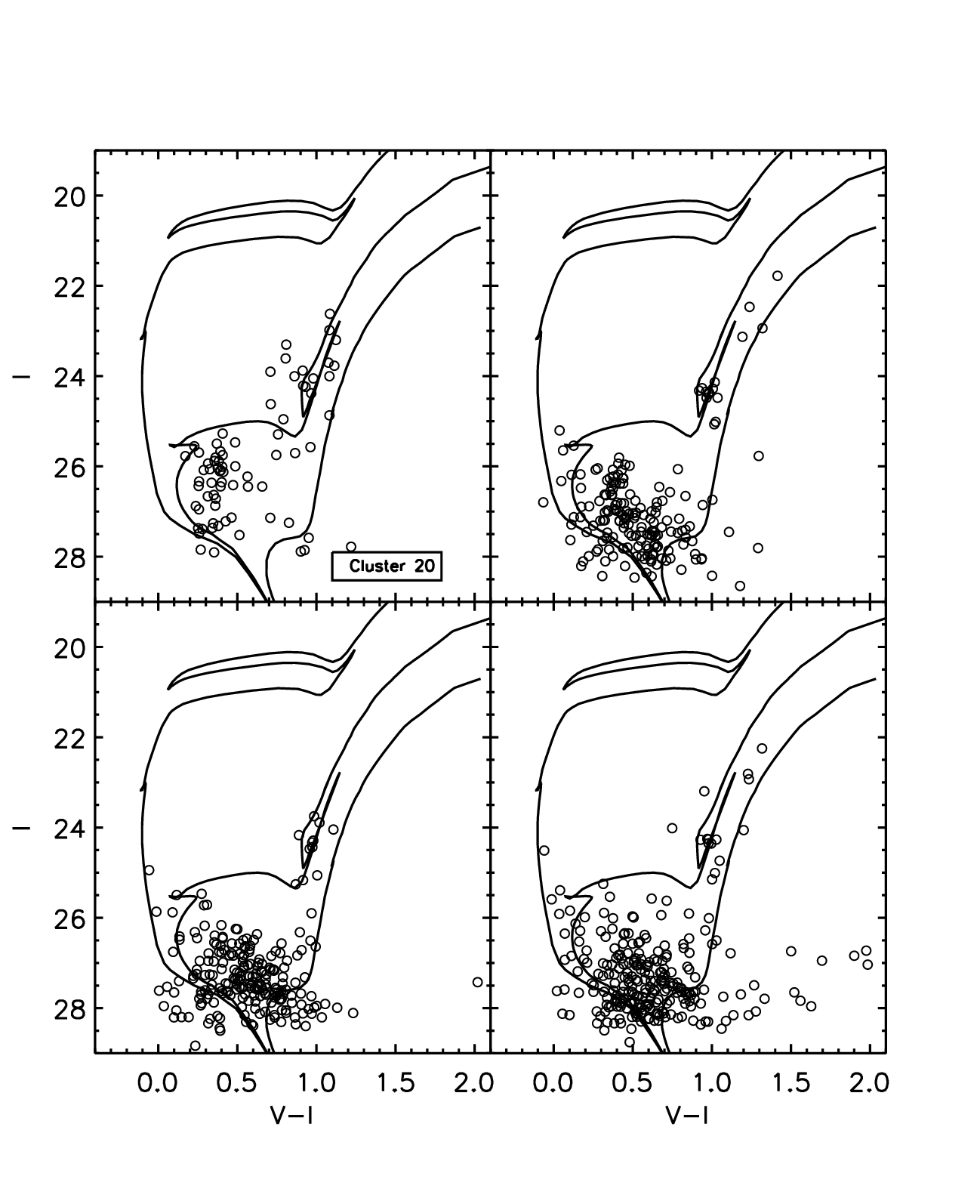}
\hfil
\includegraphics[bb= 87 377 660 1110,width=0.41\textwidth,clip=true]{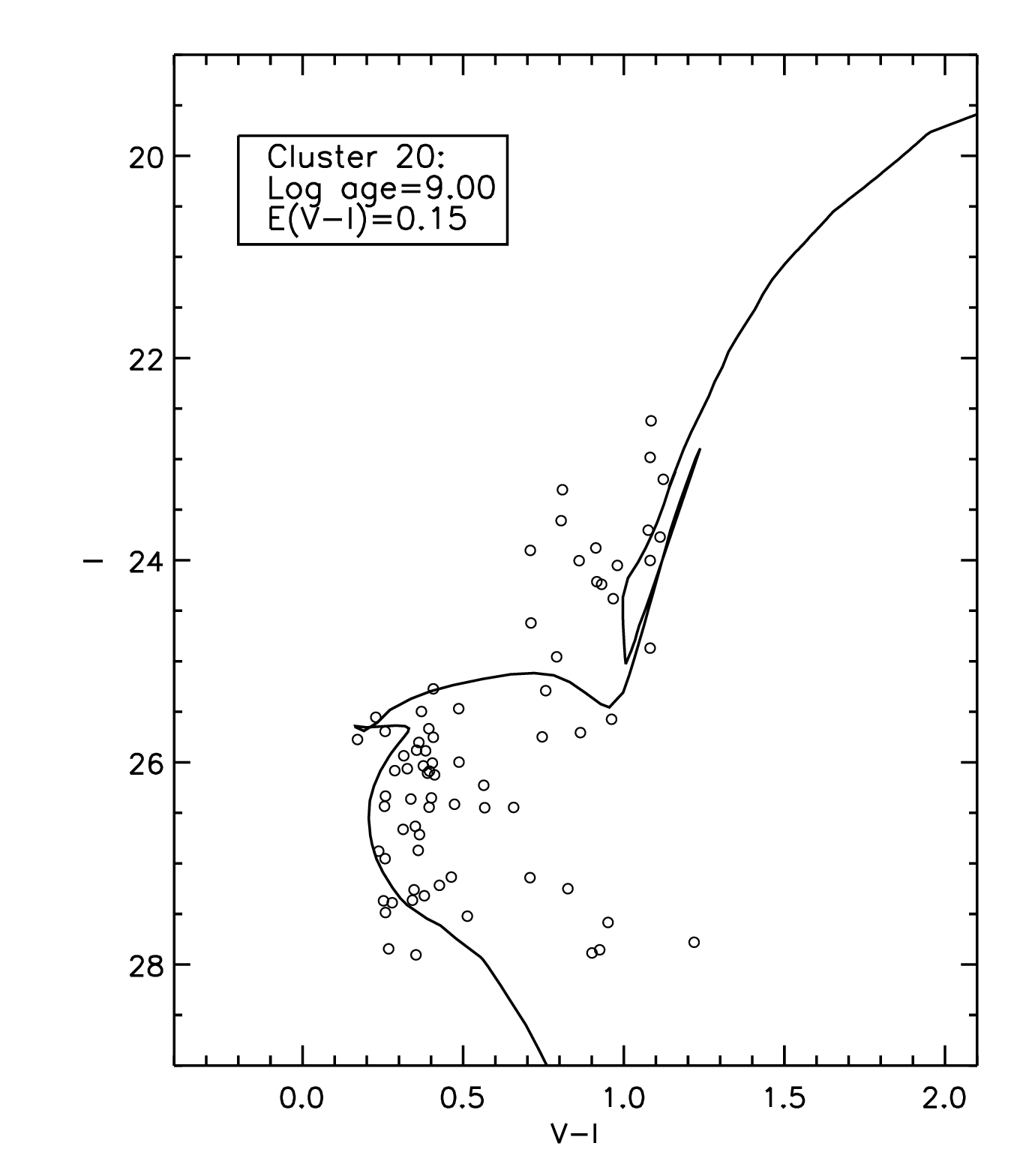}
\caption{Same as Fig. \ref{cmd1}, but for clusters 20.}
\label{cmd4}
\end{center}
\end{figure}

\begin{figure}
\begin{center}
\epsscale{0.9}
\includegraphics[bb= 59 409 622 1038,width=0.41\textwidth,clip=true]{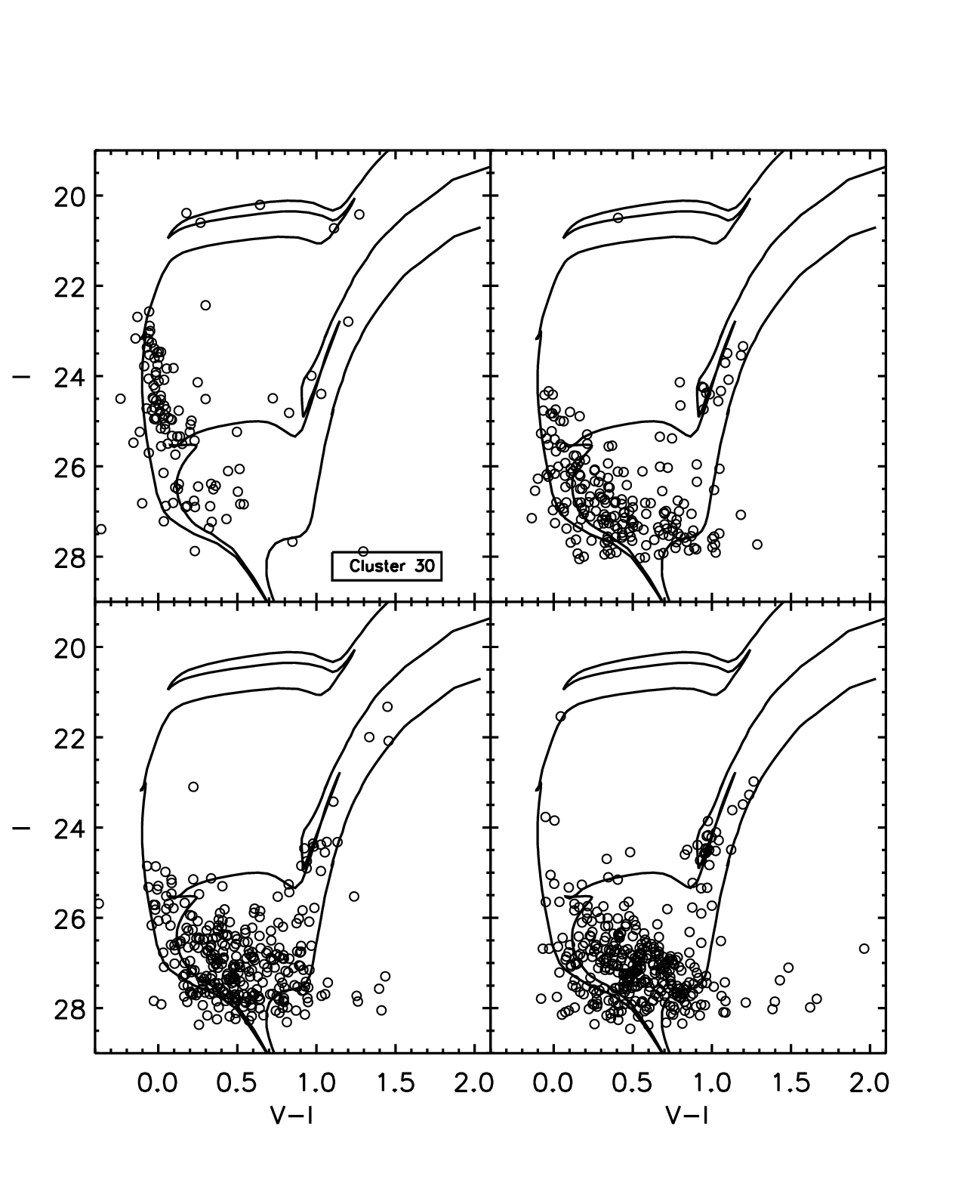}
\hfil
\includegraphics[bb=87 377 660 1110,width=0.41\textwidth,clip=true]{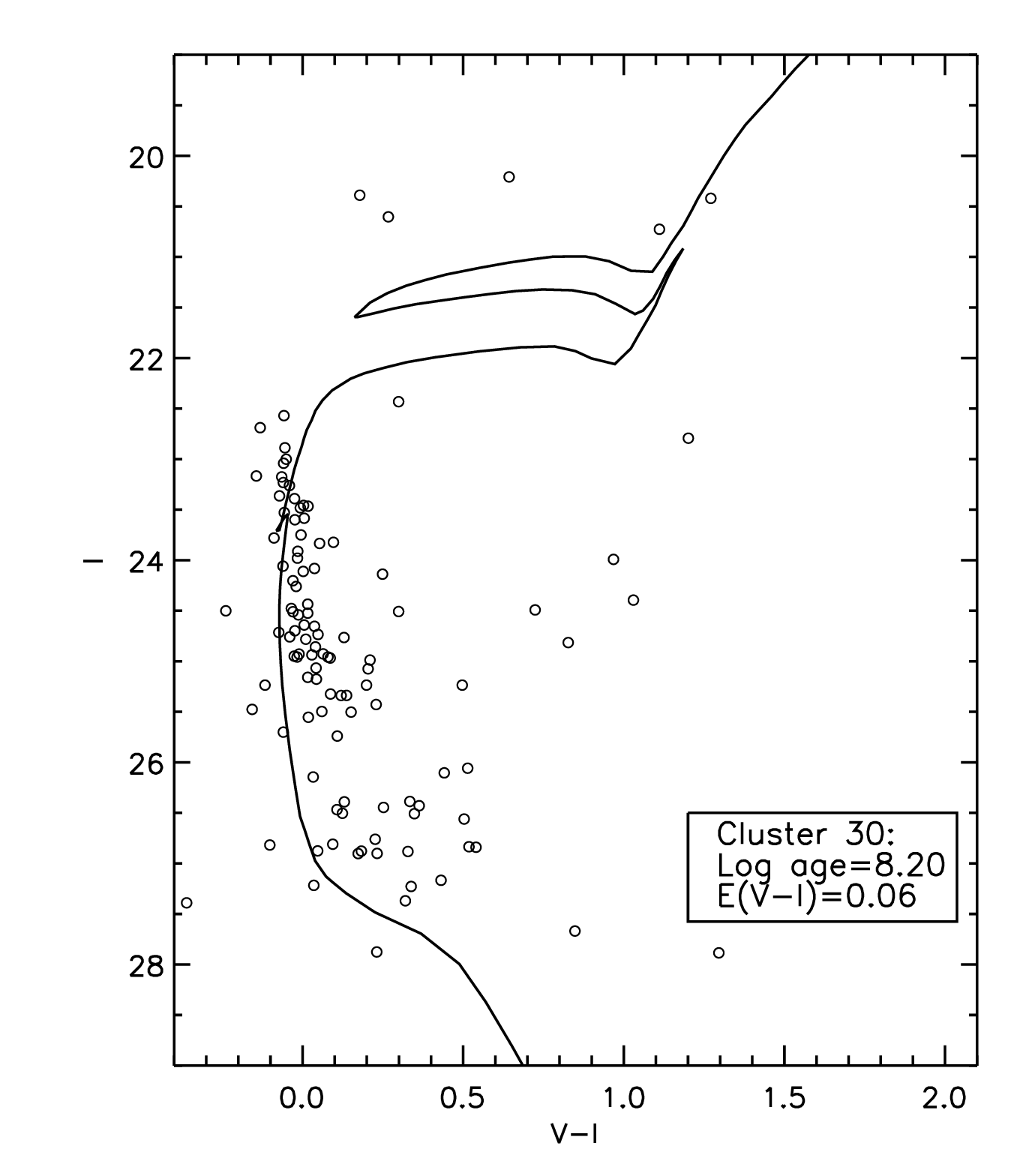}
\caption{Same as Fig. \ref{cmd1}, but for cluster 30.}
\label{cmd5}
\end{center}
\end{figure}

\begin{figure}
\begin{center}
\epsscale{0.9}
\includegraphics[bb= 59 409 622 1038,width=0.41\textwidth,clip=true]{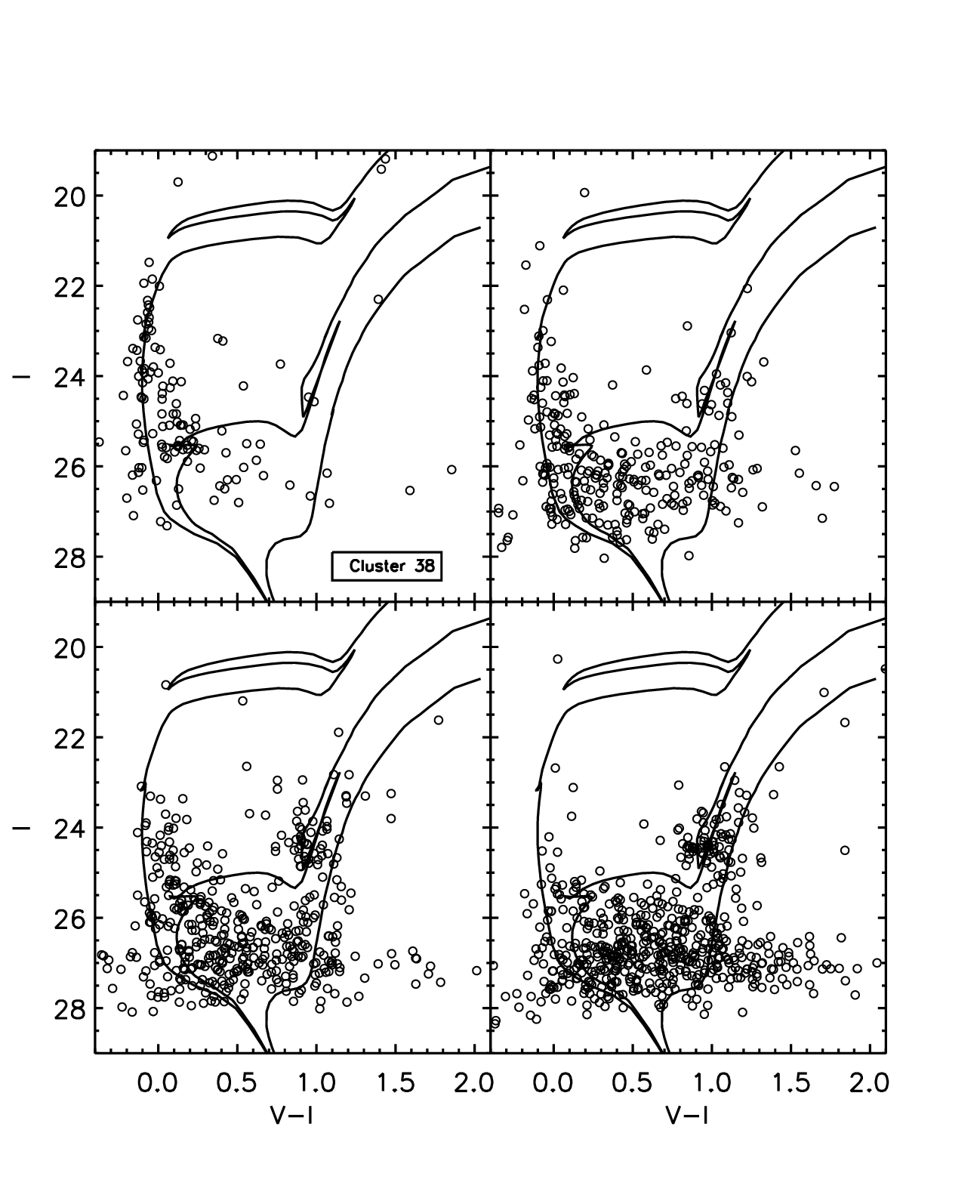}
\hfil
\includegraphics[bb=87 377 660 1110,width=0.41\textwidth,clip=true]{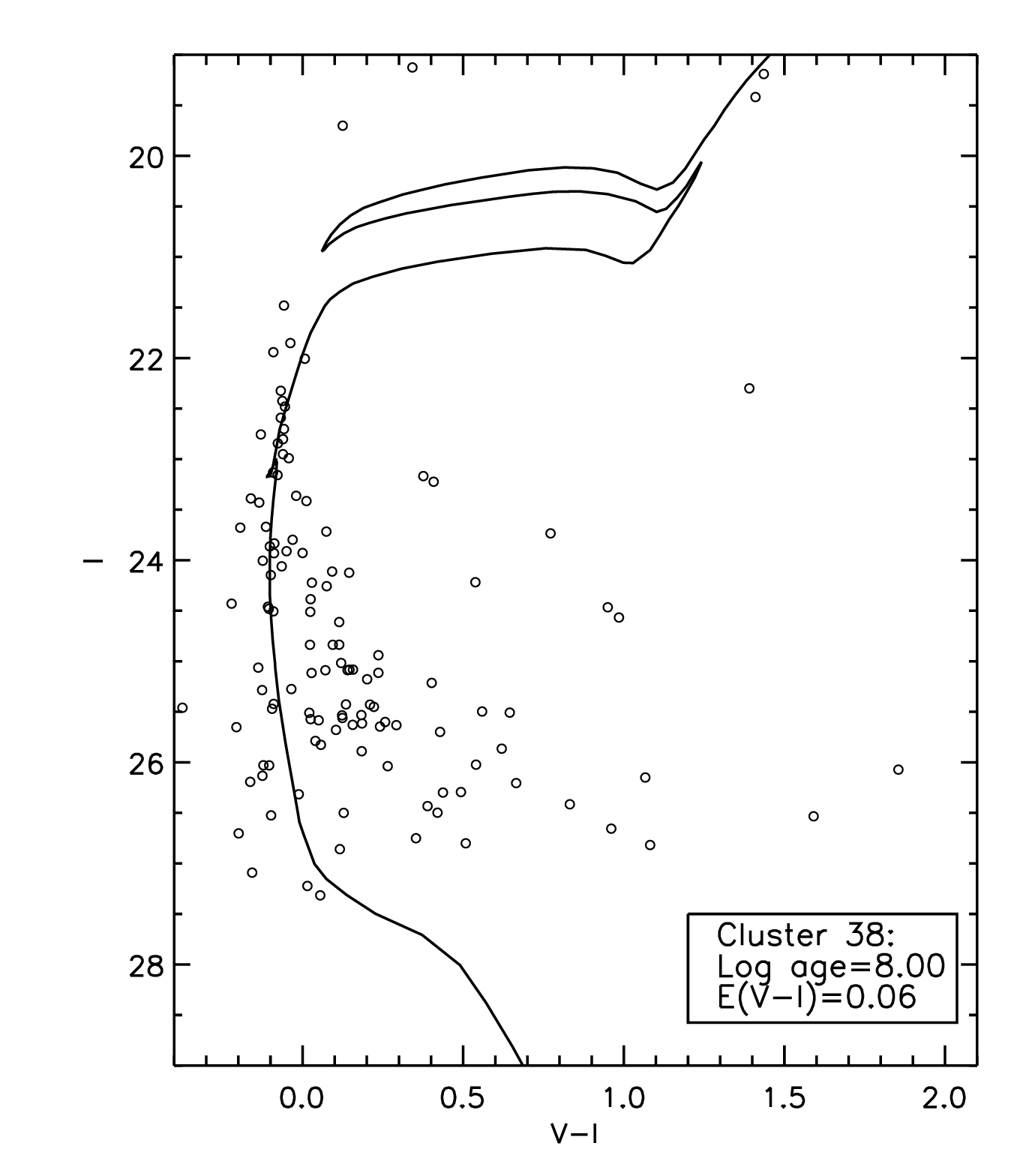}
\caption{Same as Fig. \ref{cmd1}, but for cluster 38.}
\label{cmd5}
\end{center}
\end{figure}

\begin{figure}
\begin{center}
\epsscale{0.9}
\includegraphics[bb= 59 409 622 1038,width=0.41\textwidth,clip=true]{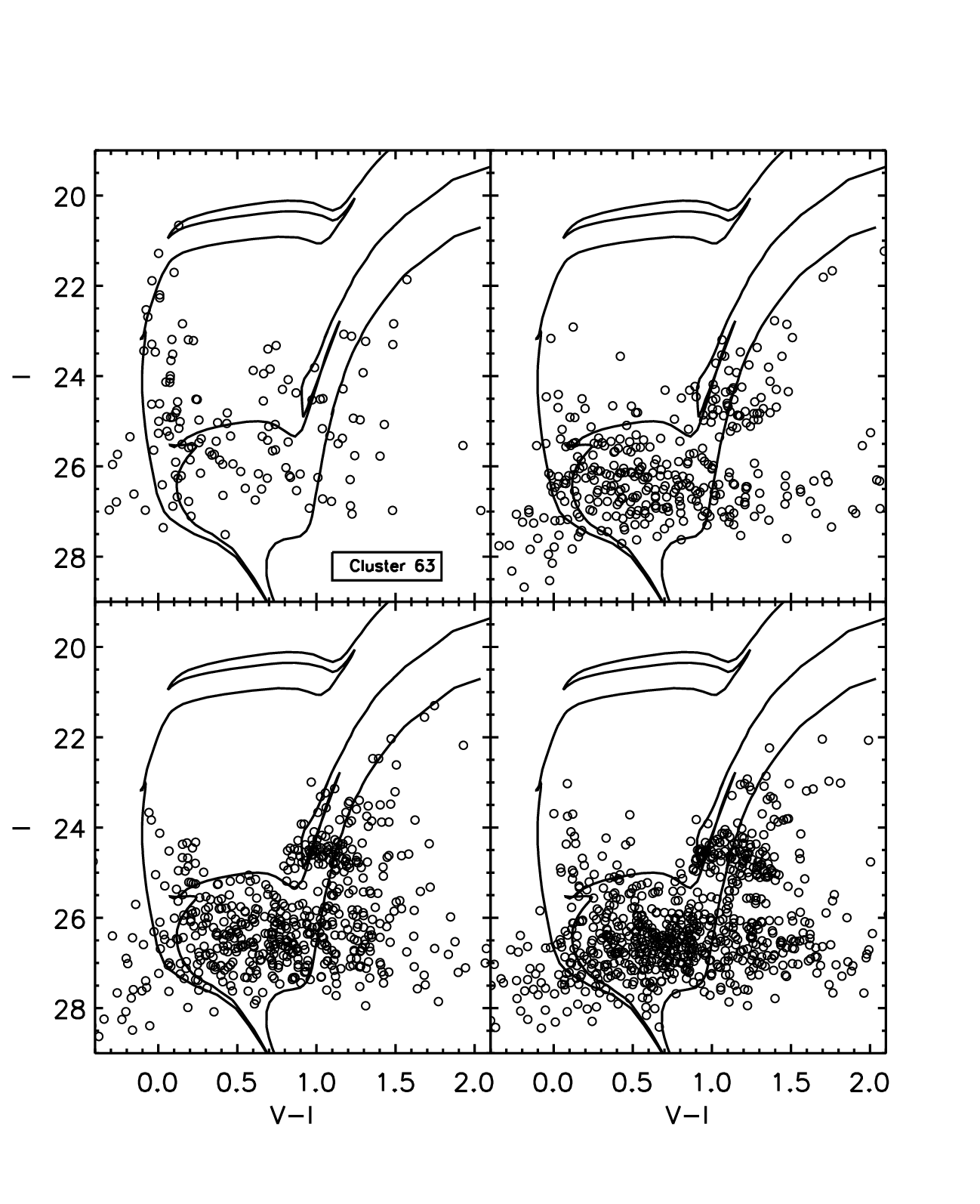}
\hfil
\includegraphics[bb=87 377 660 1110,width=0.41\textwidth,clip=true]{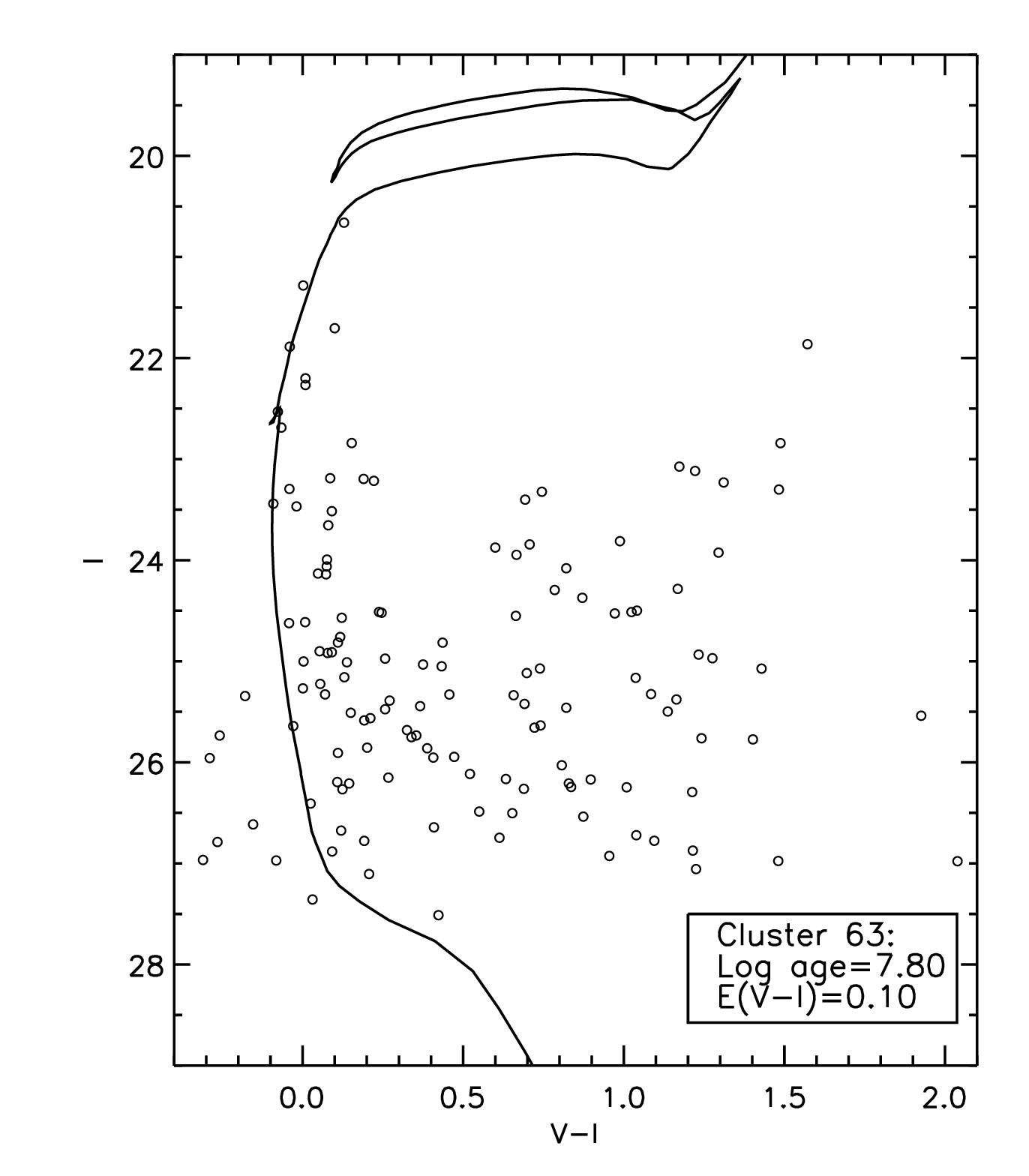}
\caption{Same as Fig. \ref{cmd1}, but for cluster 63.}
\label{cmd5}
\end{center}
\end{figure}



%




\end{document}